\documentclass[useAMS,usedcolumn,usenatbib,usegraphicx]{mn2e}

\usepackage{epsfig,graphicx,natbib}
\usepackage{./reference/mycite}
\usepackage{amssymb}
\usepackage{gensymb}
\usepackage{amsfonts}
\usepackage{amsmath}
\usepackage{color}
\usepackage{lscape,longtable}
 
\begin{document}

\title[]{Heavily Reddened Quasars at $z\sim$2 in the UKIDSS Large Area Survey: A Transitional Phase in AGN Evolution} 
\author[M. Banerji et al.]{ \parbox{\textwidth}{
Manda Banerji$^{1}$\thanks{E-mail: mbanerji@ast.cam.ac.uk}, Richard G. McMahon$^{1,2}$, Paul C. Hewett$^{1}$, Susannah Alaghband-Zadeh$^{1}$, Eduardo Gonzalez-Solares$^{1}$ \& Bram P. Venemans$^{3}$.   
}
  \vspace*{6pt} \\
$^{1}$Institute of Astronomy, University of Cambridge, Madingley Road, Cambridge, CB3 0HA, UK.\\
$^{2}$Kavli Institute for Cosmology, University of Cambridge, Madingley Road, Cambridge, CB3 0HA, UK.\\
$^{3}$Max-Planck-Institut fur Astronomie, Konigstuhl 17, D-69117, Heidelberg, Germany \\
} 

\maketitle

\begin{abstract} 

We present a new sample of purely near infra-red selected $K_{\rm{Vega}}<$16.5 [$K_{\rm{AB}}<$18.4] extremely red ($(J-K)_{\rm{Vega}}>2.5$)
quasar candidates at $z\sim$2 from $\simeq$900\,deg$^2$ of data in the UKIDSS Large Area Survey (LAS). 
Five of these are spectroscopically confirmed to 
be heavily reddened Type 1 AGN with broad emission lines bringing our total sample of reddened quasars from the UKIDSS-LAS
to 12 at $z$=1.4--2.7. At these redshifts, H$\alpha$ (6563\AA\@) is in the $K$-band. 
However, the mean H$\alpha$ equivalent width of the reddened quasars is only ten per cent larger than that 
of the optically selected population and cannot explain the extreme colours. Instead, dust 
extinction of A$_V \sim$2--6 mags is
required to reproduce the continuum colours of our sources. This is comparable to the dust extinctions 
seen in submillimetre galaxies at similar redshifts. We argue that the AGN are likely being observed
in a relatively short-lived \textit{breakout} phase when they are expelling gas and dust following a massive starburst, subsequently turning into 
UV-luminous quasars. Some of our quasars show direct evidence for strong outflows (v$\sim$800--1000 km/s) affecting the H$\alpha$ line consistent 
with this scenario. We predict that a larger fraction of reddened quasar hosts are likely to be submillimetre bright 
compared to the UV-luminous quasar population. We use our sample to place new constraints on the fraction of obscured Type 1 AGN likely to be missed in optical surveys. Taken at face-value our findings suggest that the obscured fraction depends on quasar luminosity. The space density of obscured quasars is $\sim$5$\times$ that inferred for UV-bright quasars from the SDSS luminosity function at M$_i < -30$ but seems to drop at lower luminosities even accounting for various sources of incompleteness in our sample. We find that at M$_i \sim -28$ for example, this fraction is unlikely to be larger than $\sim$20 per cent although these fractions are highly uncertain at present due to the small size of our sample. A deeper $K$-band survey for highly obscured quasars is clearly needed to test this hypothesis fully and is now becoming possible with new sensitive all-sky infra-red surveys such as the VISTA Hemisphere Survey and the \textit{WISE} All Sky Survey.

\end{abstract}


\section{INTRODUCTION}

The last few years have seen an explosion in multi-wavelength surveys
covering very large areas of sky and extending to wavelengths from the
X-ray through the optical/infra-red and into the sub-millimetre,
millimetre and radio. Such surveys are not only allowing us to gain a
more holistic view of galaxy and active galactic nuclei (AGN)
properties by sampling their entire spectral energy distribution
(SED), but their unprecedented area also opens up discovery space for
rare and unusual classes of objects e.g. the most distant
\citep{Mortlock:11} as well as the most luminous quasars
\citep{Irwin:98}. Observational studies of distant quasars and
galaxies have gained new impetus recently with the advent of sensitive
infra-red detectors which are allowing us to observe the SEDs of galaxies
and quasars after they have been redshifted beyond optical
wavelengths. In addition to sources which appear red due to their high
redshifts, there is also a population of red sources
where the red colours can be attributed to: an intrinsically red
continuum; significant dust extinction associated with the quasar host
galaxy; the presence of broad absorption lines; or a gravitationally
lensed system where the foreground lensing galaxy and/or intervening
absorption system redden the light \citep{Gregg:02, Lacy:02}. Long
wavelengths such as the far infra-red and sub-millimetre have for some
time been used to track very red, dust enshrouded populations of
galaxies \citep{Blain:02, Smail:02} that are otherwise obscured in the
optical. Now, the emergence of very large area surveys in the near and
mid infra-red wavelengths such as the UKIDSS Large Area Survey (LAS;
\citet{Lawrence:07}), VISTA Hemisphere Survey (VHS; McMahon et al. in
preparation) and Wide Infra-red Survey Explorer (\textit{WISE};
\citet{Wright:10}), are providing a different perspective on studies
of red populations of galaxies and quasars. While these surveys are
relatively shallow, the huge area of sky surveyed allows detection of
rare, luminous sources.

Quasar selection in large surveys has typically relied on detecting
their excess in flux relative to stars at short UV wavelengths
\citep{Richards:02}. However, quasars also display an infra-red excess
in the $K$-band \citep{Warren:00} which allows detection of a
complementary sample of infra-red bright sources that are relatively
faint at optical wavelengths \citep{Maddox:08}. Although numerous
attempts have previously been made at studying the infra-red bright
quasar population, the prevalence of Galactic stars with very similar
colours to red quasars in near infra-red (NIR) surveys has meant that
the selection of these objects has typically had to rely on
optical-NIR colour cuts \citep{Maddox:08} or matching to radio surveys
\citep{Glikman:07}. A consequence of the heterogenous selection
criteria used to isolate red quasars is that the exact fraction of
obscured quasars, that are not detected in optical surveys,
remains unclear. The estimates currently range from $<$20 per
cent \citep{Richards:03, Maddox:08} to 60 per cent
\citep{Glikman:07}. As dust-obscured quasars are expected to be bright at near infra-red wavelengths, a homogenous flux-limited sample selected in the NIR allows one to estimate their true space density without requiring matches to other multi-wavelength surveys which inevitably introduce additional selection biases on the sample.

In Hawthorn et al. (2012, In Prep; hereafter Paper I), we introduce a
technique for isolating stellar Extremely Red Objects (EROs) from large area
NIR-surveys using a $(J-K)$ colour selection. It is shown that, provided the artifacts that dominate
colour selected samples of extreme objects are well understood,
selecting bright EROs with stellar morphologies could prove effective
in assembling a homogenous population of infra-red selected dusty Type
1 AGN. Spectroscopic follow-up of ten sources over $\sim$100\,deg$^2$
in Paper I has led to confirmation that seven are extremely red Type 1 AGN
with similar H$\alpha$ equivalent widths to samples selected in the
optical. The red colours in these quasars imply dust extinction
values of A$_V\sim$2.3.

In the current work, we present new spectra from a search for stellar EROs
over a larger area using the same techniques as in Paper I. Taking the
samples from both papers, we focus on the inferred properties of our
infra-red bright Type 1 AGN and compare them to quasars selected at
other wavelengths; from the optical to the
sub-millimetre. \citet{Sanders:88} was the first to postulate that
sub-millimetre luminous galaxies (SMGs) and quasars are different
observational manifestations of the same sources. Major merger induced
starbursts appear as sub-millimetre galaxies and as the dust clears
from the decaying starburst, the central nuclear region is revealed as
an optically bright quasar. Red broad absorption-line (BAL) quasars that are enshrouded by dust have also been studied e.g. by \citet{Egami:96} and hypothesized to correspond to a young phase in quasar evolution. A sample of NIR-selected red quasars may therefore
provide crucial observational evidence for the link between galaxies
and supermassive black holes, representing a short-lived phase in the
evolution of the galaxy where the dust has not yet fully cleared and
is still obscuring the view of the quasar in the optical wavelengths.

Throughout this paper we assume a
flat concordance cosmology with $\rm H_0$=70 km s$^{-1}$ Mpc$^{-1}$, 
$\rm \Omega_M$=0.3 and $\Omega_\Lambda$=0.7. All magnitudes quoted are in
the native system of the survey being discussed - i.e. $AB$ for SDSS
and Vega for UKIDSS unless otherwise stated. The $AB$-Vega conversions
for both the SDSS and UKIDSS filters can be found in
\citet{Hewett:06}.

%The AB-Vega conversions for both the SDSS and UKIDSS filters can be
%found in \citep{Hewett:06}. The Vega-AB conversions for the WISE
%filters are +2.683, +3.319, +5.242 and +6.404 for the 3.4,4.6,12 and
%22$\mu$m bands respectively.

\section{DATA}

\label{sec:data}

Red quasar candidates are selected from the UKIDSS LAS, one of the
largest NIR surveys and the NIR counterpart to the Sloan Digital Sky
Survey in the northern hemisphere. In Paper I, we have presented
details of our candidate selection algorithm including how to
effectively screen the data for the many spurious sources. Below, we
provide a brief overview of some of the key features of our selection
but the interested reader is referred to Paper I for more information.

\subsection{Photometric Selection}

\label{sec:photom}

\begin{figure*}
\begin{center}
\includegraphics[scale=0.5,angle=0]{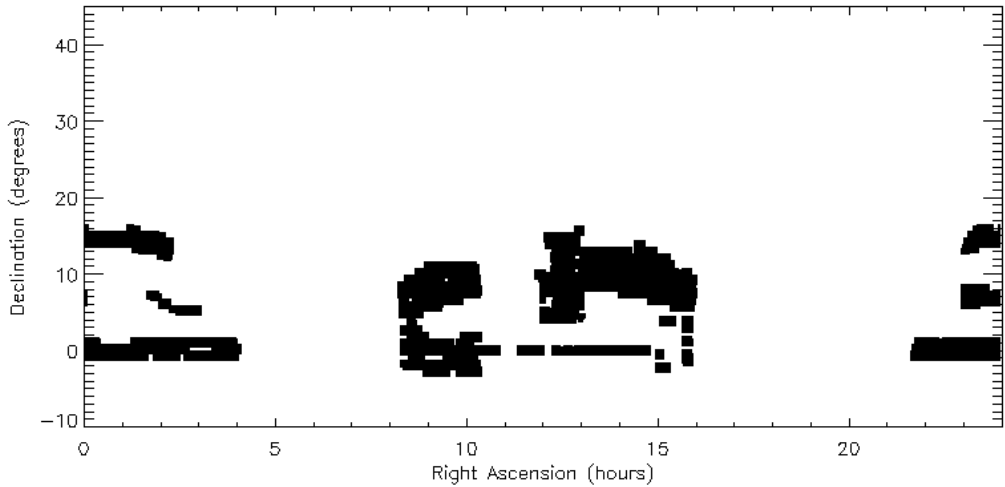}
\caption{The UKIDSS-LAS DR4 footprint from which all targets in this paper 
have been selected.}
\label{fig:ulasdr4}
\end{center}
\end{figure*}

Red quasar candidates were selected from the UKIDSS-LAS Data Release 4
(DR4) by isolating sources with stellar morphologies and
J$-$K$_{\rm{Vega}}>$2.5 [$J-K_{AB}>$1.5] and $K_{\rm{Vega}}<$16.5
[$K_{AB}<$18.4]. They belong to a class of sources known as Extremely
Red Objects (EROs). The UKIDSS-LAS DR4 footprint is shown in
Fig.~\ref{fig:ulasdr4}. Two samples were defined: \ 1)
08\,h$<$RA$<$16\,h and 0$\degree$$<$DEC$<$15$\degree$ where we require
a non-detection in the optical SDSS bands in order for the source to
be included in our spectroscopic sample, and 2) the SDSS southern
stripe Stripe82 region at 21\,h$<$RA$<$00\,h and
$-$1.25$\degree$$<$DEC$<$1.25$\degree$ which benefits from deep co-added
optical data \citep{Annis:11}. In the case of the Stripe82 sample, we use the deep
co-add images to select sources that are fainter than $i_{AB}$=20.5
and redder than $(i-K)_{AB}$=2.5 ($(i-K)_{\rm{Vega}}$=4.4) as was the case for the North
Galactic Pole (NGP) part of the Stripe82 sample presented in Paper
I. We also include four sources that are brighter than this $i$-band
limit in order to check whether there are any reddened AGN that could
potentially be missed by applying the SDSS $i$-band
selection. Finally, to assess the utility of X-ray detections for
identifying red, luminous AGNs, two objects included in the 2XMM X-ray
catalogue \citep{Watson:09} that satisfy $K_{\rm{Vega}}<$17.0
[$K_{AB}<$18.9] were also targeted. The objects are the two brightest
EROs that are known X-ray sources in the survey area.

The full area spanned by Sample 1 is 770\,deg$^2$, calculated using
the publicly available \textsc{Mangle} software
\citep{Swanson:08}. There are 43 EROs with $K_{\rm{Vega}}<$16.5 and
$(J-K)_{\rm{Vega}}>$2.5. Nine of the 43 sources are undetected in the
SDSS and therefore are candidate reddened Type 1 AGN. In addition, ULASJ0908+1042 was targetted despite being detected in SDSS as it is also a FIRST radio source. Finally, ULASJ1002+0137 was also targetted despite being fainter in $K$ as it is an X-ray source. Sample 1 therefore comprises 11 red quasar candidates which are summarised in Table \ref{tab:sample}.

The South Galactic Pole (SGP) part of Sample 2 that overlaps the SDSS
Stripe82 spans an additional 100\,deg$^2$. There are 14 EROs in this
area with $K$(Vega)$<$16.5 and $(J-K)_{\rm{Vega}}>$2.5. Only three of
these are significantly fainter than $i$=20.5 in the deeper SDSS
co-add images and redder than $(i-K)_{AB}$=2.5. Two of these,
ULASJ2200+0056 and ULASJ2224$-$0015, were targetted for spectroscopy but the third,
ULASJ2346$-$0038, was unfortunately incorrectly excluded from the
spectroscopic target list as it had a close neighbour in a different
frameset and was erroneously assumed to be a duplicate source. In
addition, four other bright sources with $K_{Vega}<$16 that have
$i_{AB}<=$20.5 were followed-up in order to assess whether any
luminous $K$-band sources that are also detected in SDSS could
potentially be reddened Type 1 AGN. Finally, the fainter source ULASJ2343+0017 was also targetted on account of being detected in the X-ray. Therefore Sample 2 comprises eight potential red quasar candidates four of which have i$<=$20.5. 

\begin{figure}
\begin{center}
\includegraphics[scale=0.5,angle=0]{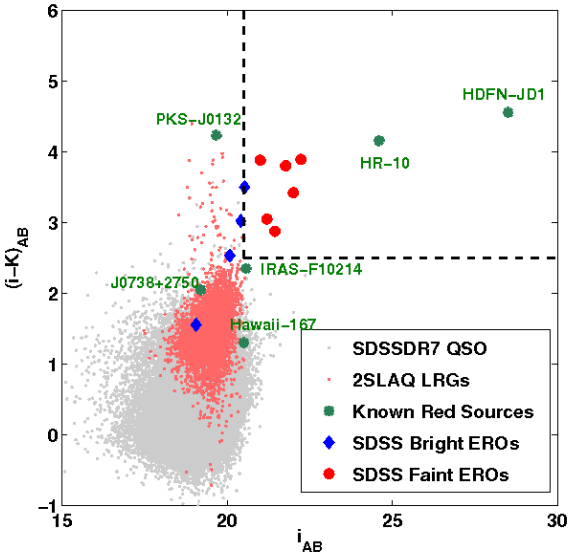}
\caption{$i-K$ vs $i_{AB}$ colour-magnitude diagram showing the EROs from
our survey overlapping the deep SDSS Stripe82 co-add data. We also
show the SDSS DR7 quasars \citep{Schneider:10} as the grey points,
Luminous Red Galaxies (LRGs) from the 2SLAQ survey \citep{Cannon:06}
as the small red points as well as various well-known red sources in
green. Our selection criteria that divides SDSS-bright and faint EROs
is shown by the dashed lines. The SDSS-bright EROs with i$<$20.5
(blue) have no lines in the NIR spectra consistent with these being
LRGs whereas the SDSS-faint EROs (red) are all spectroscopically
confirmed to be reddened Type 1 AGN.}
\label{fig:colour}
\end{center}
\end{figure}

In Table~\ref{tab:sample} we provide details of all our red quasar
candidates in both samples including those without spectra.
Fig.~\ref{fig:colour} shows the $(i-K)$ distribution of the EROs in Stripe82 with deep optical photometry. These include the SGP part of the sample
presented in Table \ref{tab:sample} of this paper as well as the NGP
part of the sample presented in Paper I. We also plot several well-known red objects such as the dusty ultraluminous galaxies HR10 \citep{Hu:94, Dey:99} and IRAS F10214 \citep{RR:91}, well-known red lensed quasars such as J0738+2750 and PKS-J0132 \citep{Gregg:02} and red broad absorption-line quasars e.g. Hawaii-167 \citep{Cowie:94, Egami:96}. As can be seen from the figure, our ERO candidates are among the reddest infra-red bright objects currently known. As discussed in detail in the
next section, all the SDSS faint EROs are spectroscopically confirmed
to be Type 1 AGN whereas no redshifts were obtained for the SDSS
bright EROs consistent with identifications as compact red galaxies.

\subsection{Spectroscopic Follow-up} 

Spectroscopic observations of our red quasar candidates were made with
the SINFONI integral field spectrograph on the VLT \citep{SINFONI:01,
SINFONI:02} between 2009 April and July (Period83:383.A-0573(A)) with
seeing ranging from 0.8--1.4$^{\prime \prime}$ and the sample of 13 red
quasar candidates that were observed, is presented in Table
\ref{tab:sample}. SINFONI was used in noAO mode with 0.25$^{\prime \prime}$x0.25$^{\prime \prime}$ pixels
corresponding to an 8$^{\prime \prime}$x8$^{\prime \prime}$ field-of-view with the R=1500
$H$+$K$ grism yielding a resolution of 12\,\AA\@ or 185\,kms$^{-1}$
(measured from sky lines) and a dispersion of 5\,\AA\@ per pixel. The
observed wavelength range is 1.4--2.45\,$\mu$m. The targets were offset by $\pm$1.5$^{\prime \prime}$
in RA and DEC from the centre of the IFU for sky subtraction purposes, however always retaining the target 
in the field of view. The array centre therefore moved in a 3$\times$3$^{\prime \prime}$ square. The data were
reduced using standard SINFONI ESOREX pipelines which include
extraction, sky subtraction, wavelength calibration and flat fielding.

\begin{table*}
\begin{center}
\caption{Sample of red quasar candidates from the UKIDSS-LAS DR4. Two
samples are constructed at 08\,h$<$RA$<$16\,h and overlapping the SGP
part of the SDSS southern stripe Stripe82. Slightly different
selection criteria were used to select sources in these two samples as
detailed in Section~\ref{sec:photom}. Both sources that were
spectroscopically followed-up using SINFONI as well as sources with no
spectra are included for completeness. The NGP part of the Stripe82
sample was presented in Paper I as well as two of the quasars in
Sample 1 that overlap the UKIDSS-LAS Early Data Release area.}
\label{tab:sample}
\begin{tabular}{lcccccl}
\hline \hline
Name & RA & DEC & K$_{\rm{Vega}}$ & i$_{AB}$ & Redshift & Notes \\
\hline
\multicolumn{7}{c}{\textbf{Sample 1 08H$<$RA$<$16H - 11 objects}} \\
\hline
ULASJ0908+1042 & 09:08:06.81 &  +10:42:39.5 & 15.13 & 21.00 & -- & FIRST radio source \\
ULASJ0946+0016 & 09:46:28.66 &  +00:16:25.8 & 16.28 & -- & no spectrum & \\
ULASJ1002+0137\footnote{COSMOS AGN} & 10:02:11.29 &  +01:37:06.8 & 17.00 & -- & 1.595 & Faint $K$-band, 2XMM X-ray source \\
ULASJ1004+0026 & 10:04:43.68 &  +00:26:09.8 & 15.51 & -- & -- & \\
ULASJ1200+0423 & 12:00:59.45 &  +04:23:38.4 & 16.30 & -- & no spectrum & \\
ULASJ1234+0907 & 12:34:27.52 &  +09:07:54.2 & 16.15 & -- & 2.503 & \\
ULASJ1319+0009 & 13:19:10.66 &  +00:09:56.1 & 16.04 & -- & -- & UKIDSS-EDR Paper I \\
ULASJ1320+0948 & 13:20:47.95 &  +09:08:05.8 & 16.46 & -- & no spectrum & \\
ULASJ1422+1023 & 14:22:39.13 &  +10:23:34.2 & 16.46 & -- & -- & FIRST radio source \\
ULASJ1455+1230 & 14:55:21.00 &  +12:30:08.6 & 16.34 & -- & 1.460 & \\
ULASJ1539+0557 & 15:39:10.16 &  +05:57:49.7 & 15.85 & -- & 2.658 & UKIDSS-EDR Paper I \\
\hline
\multicolumn{7}{c}{\textbf{Sample 2 Stripe82 SGP 21H$<$RA$<$24H - 8 objects}} \\
\hline
ULASJ2148$-$0011 & 21:48:25.35 &  $-$00:11:32.1 & 15.49 & 20.41 & -- & Bright SDSS \\
ULASJ2150$-$0055 & 21:50:48.27 &  $-$00:55:45.8 & 15.13 & 20.53 & -- & On SDSS selection limit \\
ULASJ2200+0056 & 22:00:24.87 &  +00:56:04.8 & 15.22 & 21.00 & 2.541 & Faint SDSS \\
ULASJ2219+0036 & 22:19:30.44 &  +00:36:26.4 & 15.61 & 19.06 & -- & Bright SDSS \\
ULASJ2224$-$0015 & 22:24:09.41 &  $-$00:15:24.1 & 16.07 & 21.77 & 2.223 & Faint SDSS \\
ULASJ2251+0047 & 22:51:21.84 &  +00:47:33.4 & 15.65 & 20.08 & -- & Bright SDSS \\
ULASJ2343+0017 & 23:43:39.68 &  +00:17:56.0 & 16.68 & 22.00 & -- & Faint $K$-band, 2XMM X-ray source\\
ULASJ2346$-$0038 & 23:46:58.41 &  $-$00:38:06.4 & 15.31 & 21.60 & no spectrum & Faint SDSS \\

\hline
\end{tabular}
\end{center}
\end{table*}

All spectra were flux-calibrated using standard telluric stars
observed at similar airmass. The target spectrum is divided by the
spectrum of the calibration star and multiplied by the appropriate
blackbody, depending on the spectral type of the star, in order to get
a relative flux-calibrated spectrum. Due to the absence of absolute
spectrophotometric standards in the NIR, the relative flux-calibrated
spectra are then normalised to the UKIDSS $K$-band magnitudes of the
targets in order to provide an absolute flux calibration.

%\begin{table}
%\begin{center}
%\caption{Summary of telluric standards used for flux calibration of SINFONI spectra.}
%\label{tab:calib}
%\begin{tabular}{lccccc}
%\hline \hline
%Date & Targets & RA & DEC & Type & T$_{\rm{eff}}$(K) \\
%\hline
%2009-04-13 & 1455 & 206.8297 & -2.4437 & B2V & 22300 \\
%2009-05-13 & 1002 & 228.9384 & -14.6932 & B3V & 19000 \\
%2009-06-16 & 1234 & 186.5807 & -24.2726 & G2V & 5830 \\
%2009-07-30 & 2200,2224 & 343.2805 & -49.5609 & G0V & 5930 \\
%\hline
%\end{tabular}
%\end{center}
%\end{table}

\begin{figure*}
\begin{center}
\centering
\begin{minipage}[c]{1.00\textwidth}
\begin{tabular}{ccc}
\includegraphics[scale=0.35,angle=0]{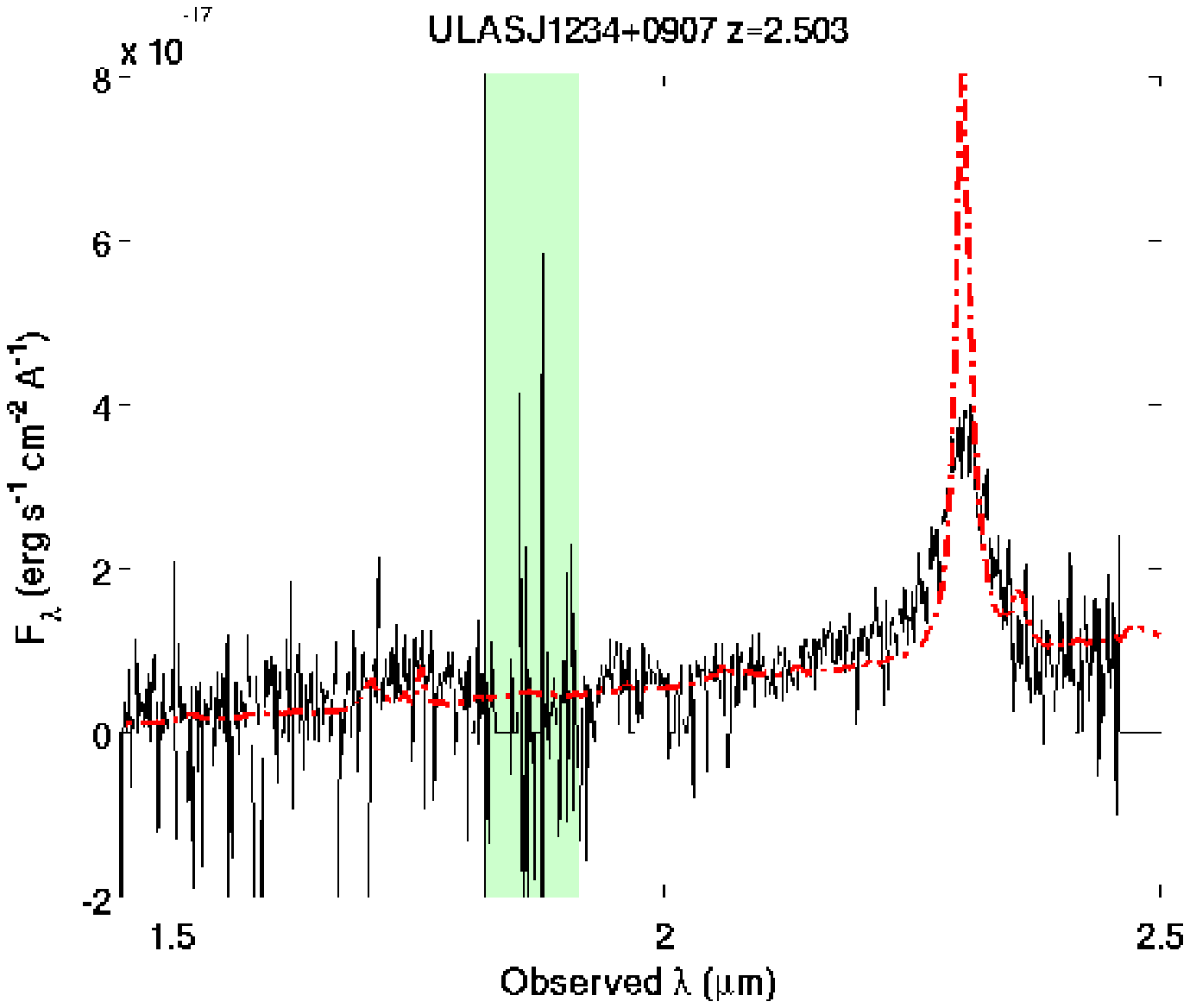} & \includegraphics[scale=0.35,angle=0]{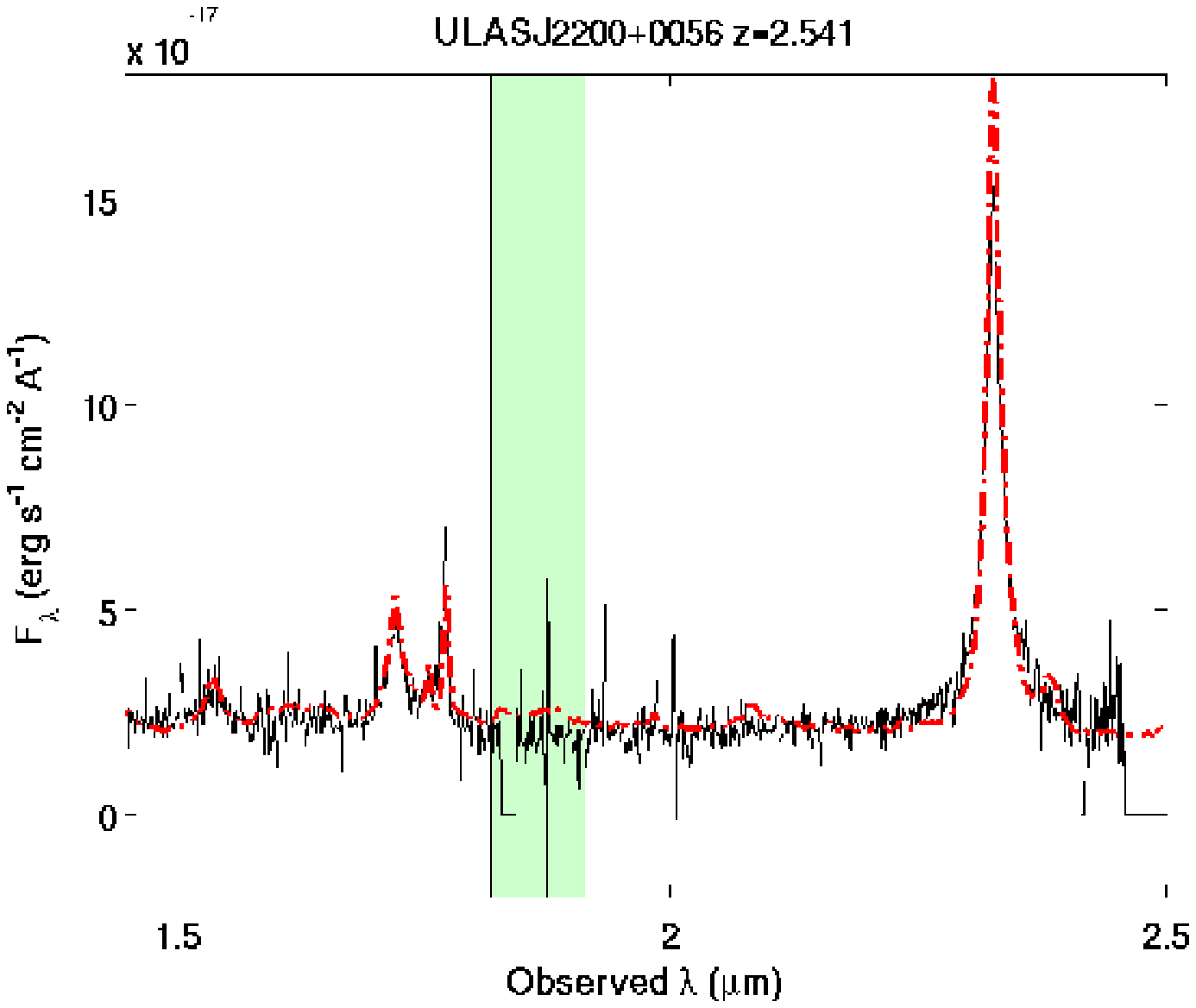} & \includegraphics[scale=0.35,angle=0]{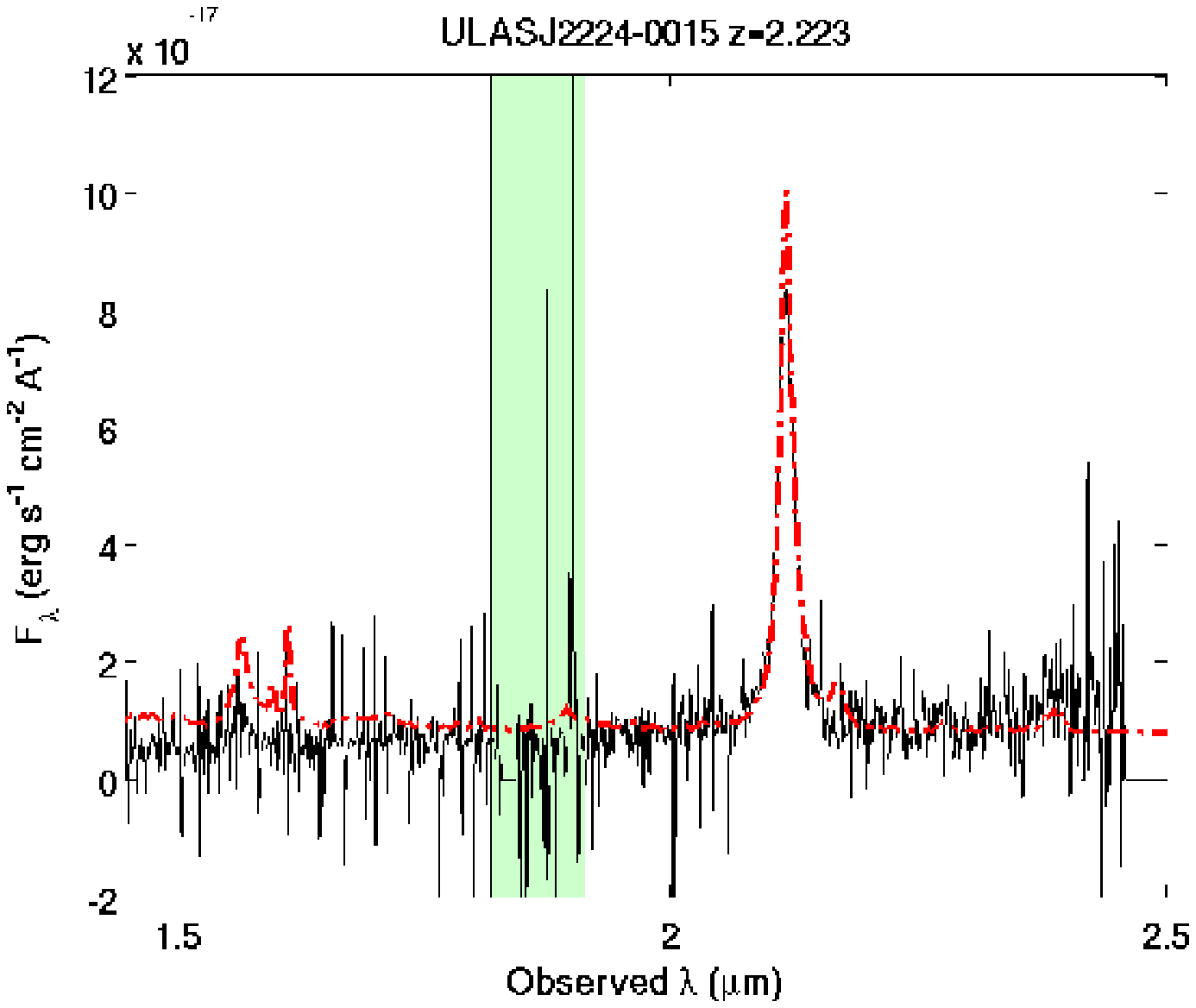} \\
\includegraphics[scale=0.35,angle=0]{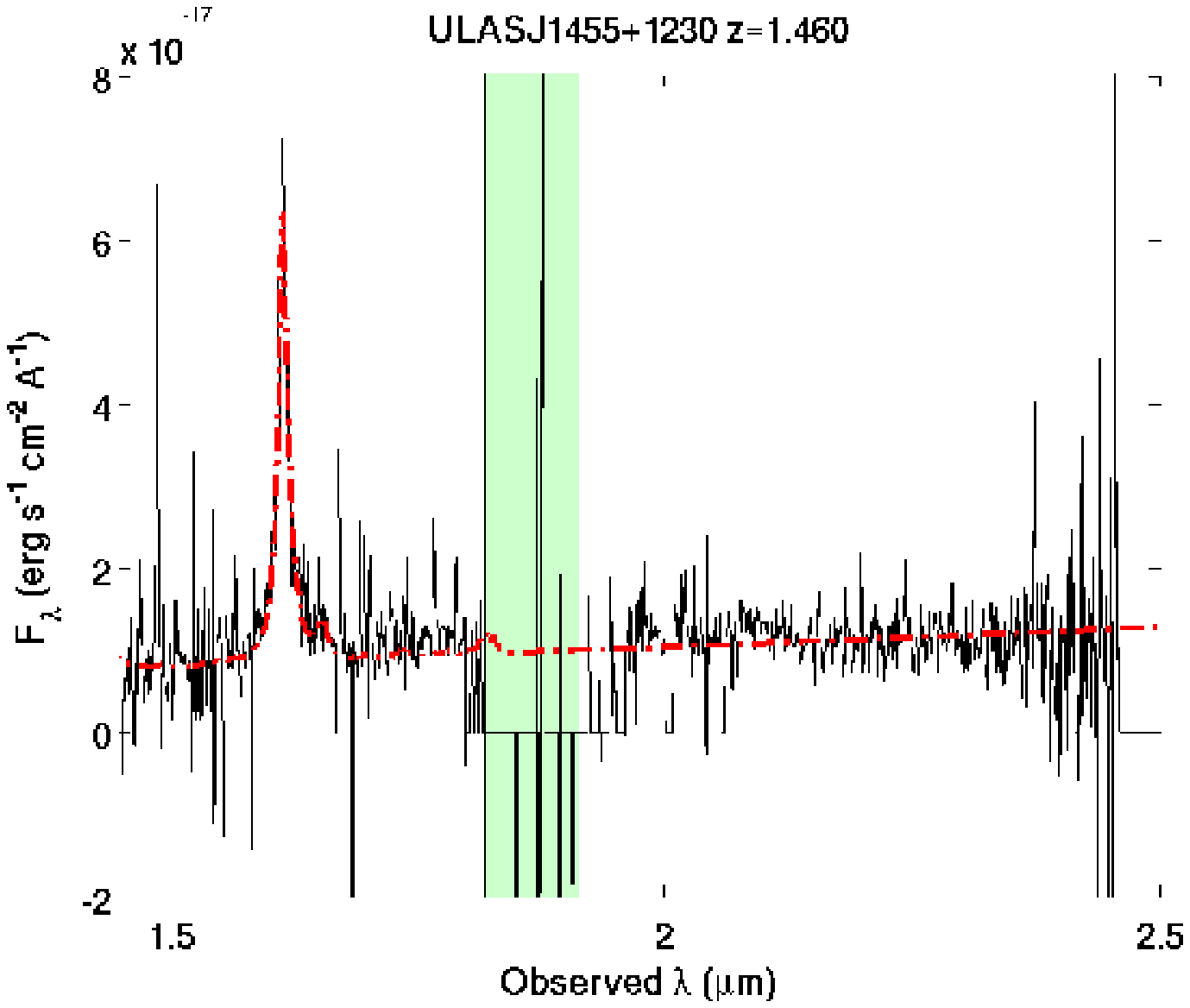} & \includegraphics[scale=0.35,angle=0]{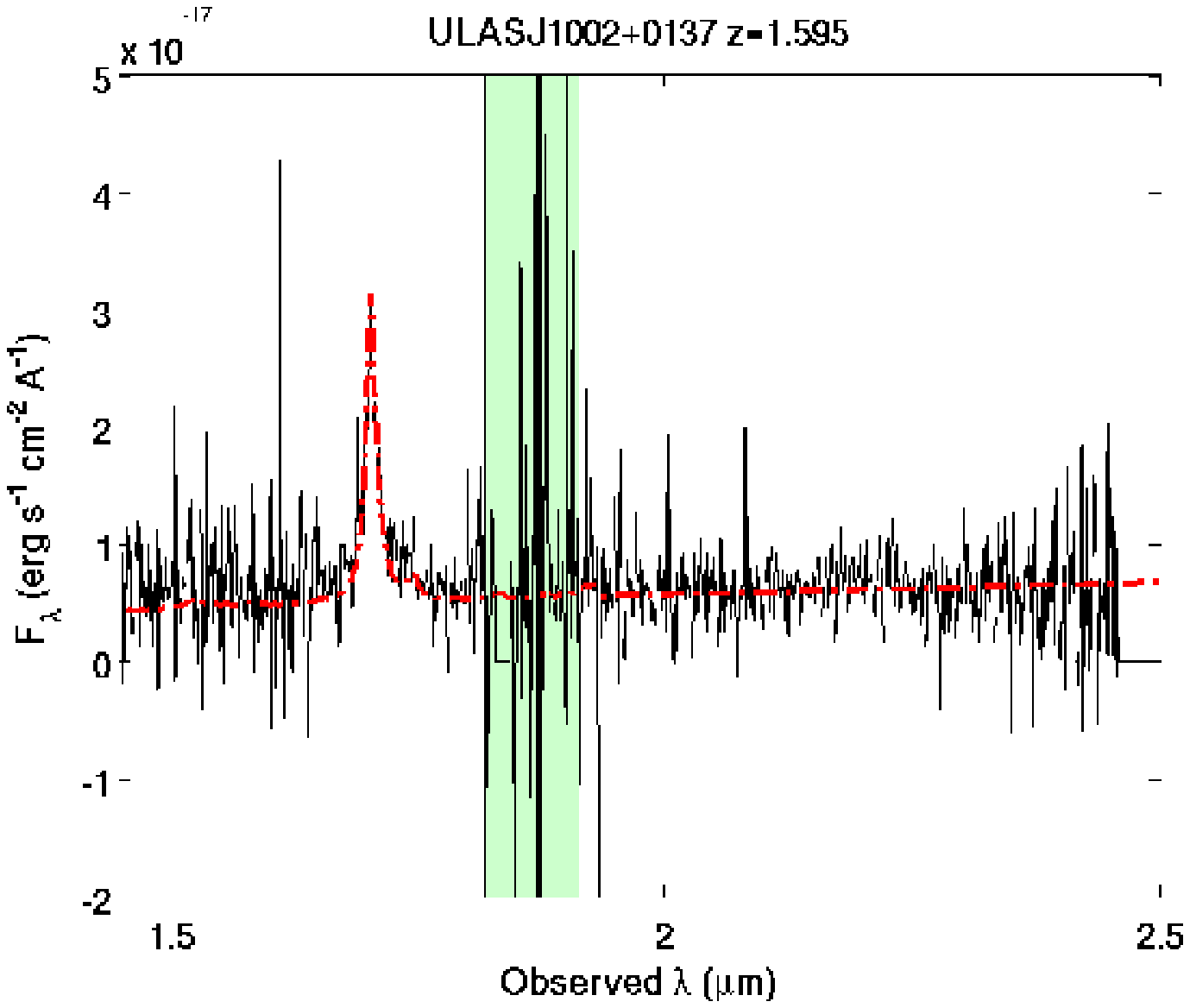} & \includegraphics[scale=0.35,angle=0]{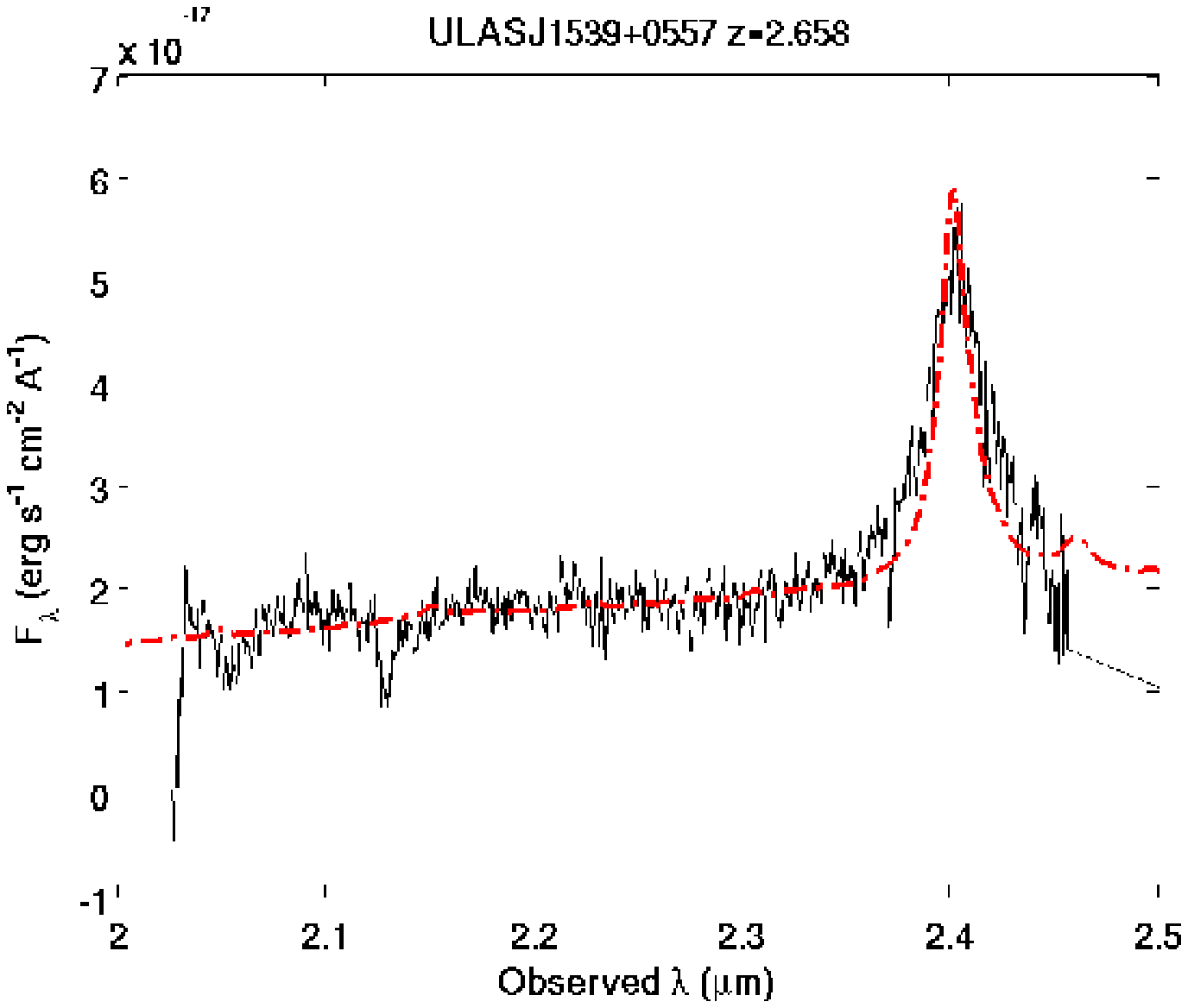} \\
\includegraphics[scale=0.35,angle=0]{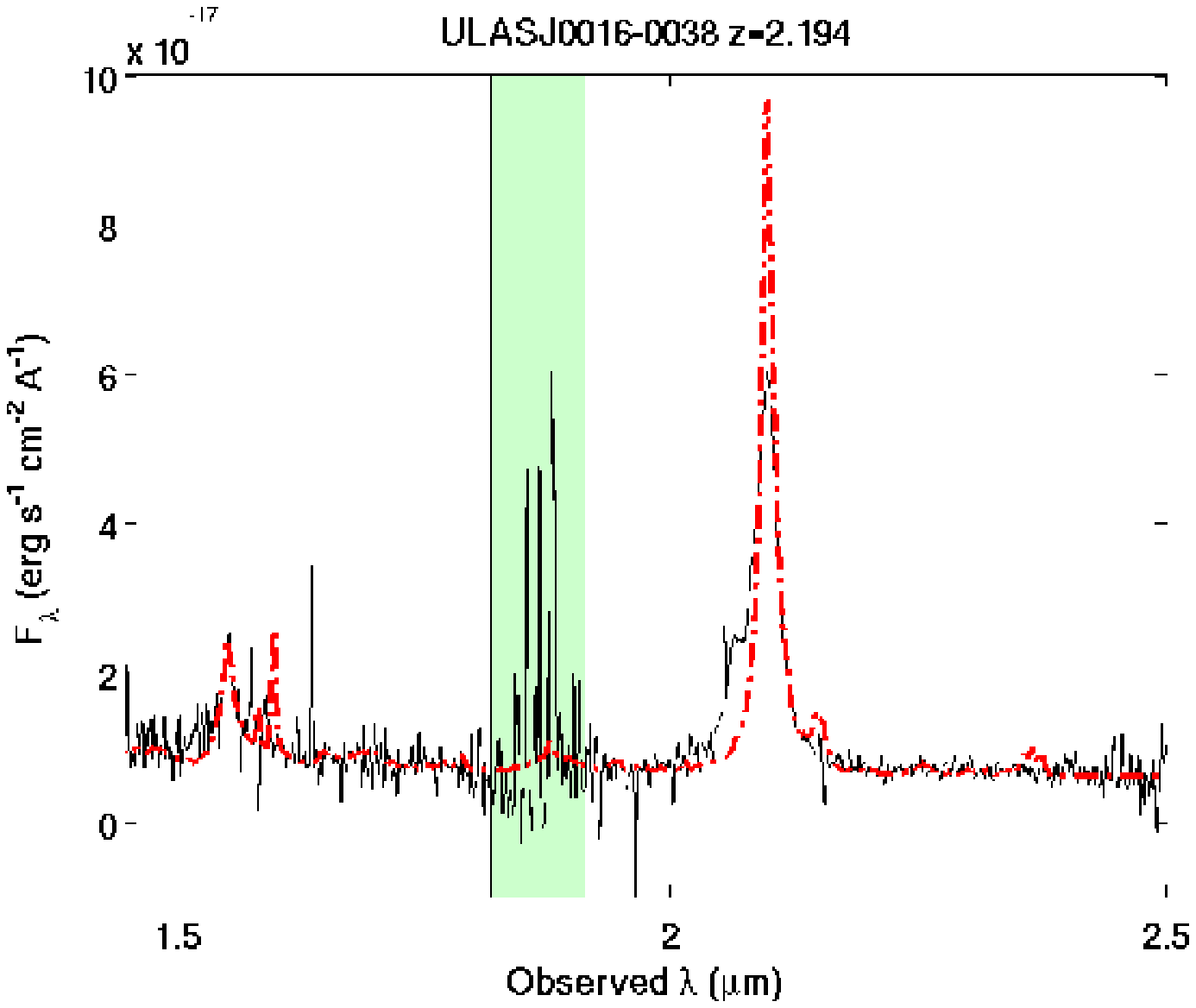} & \includegraphics[scale=0.35,angle=0]{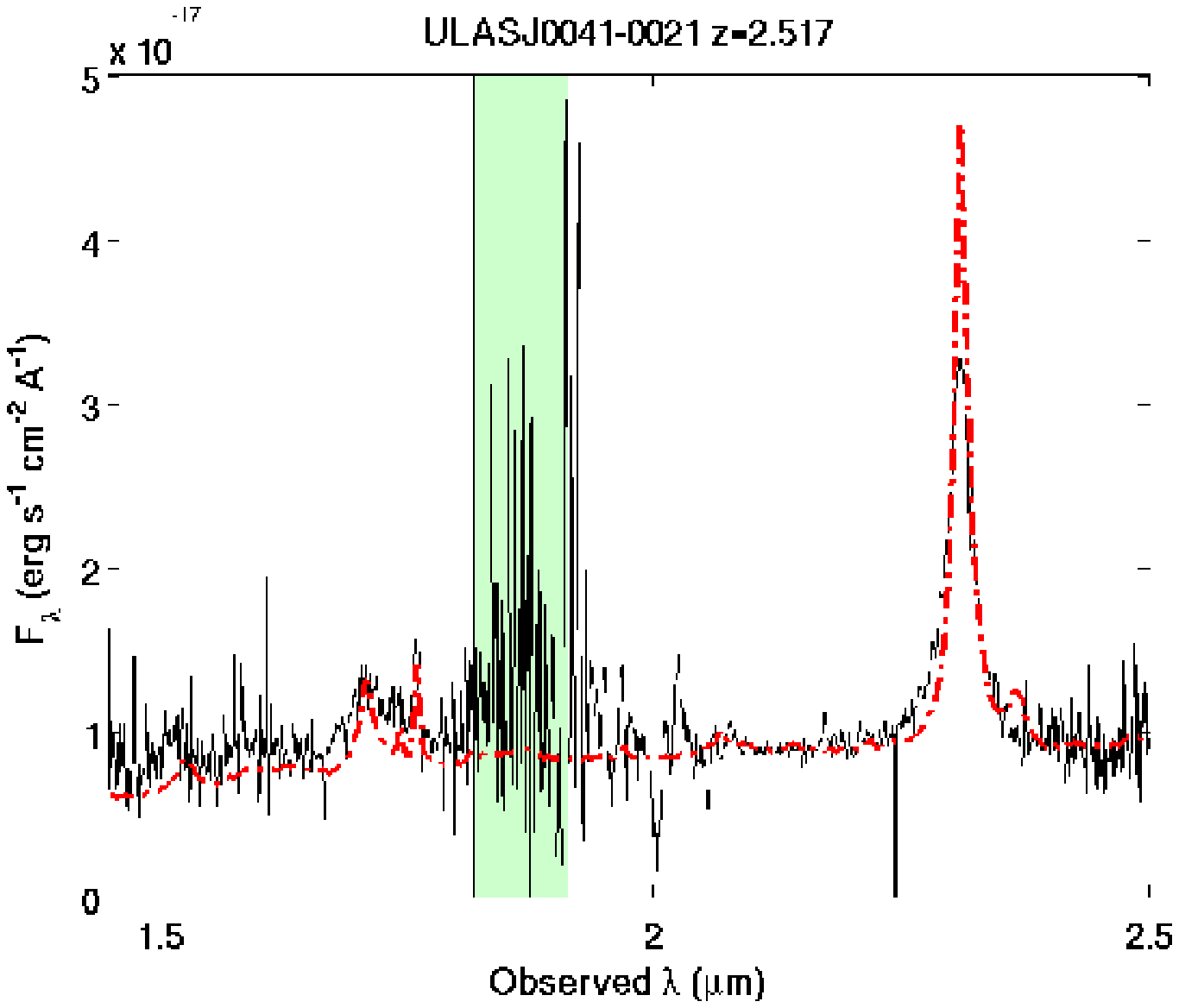} & \includegraphics[scale=0.35,angle=0]{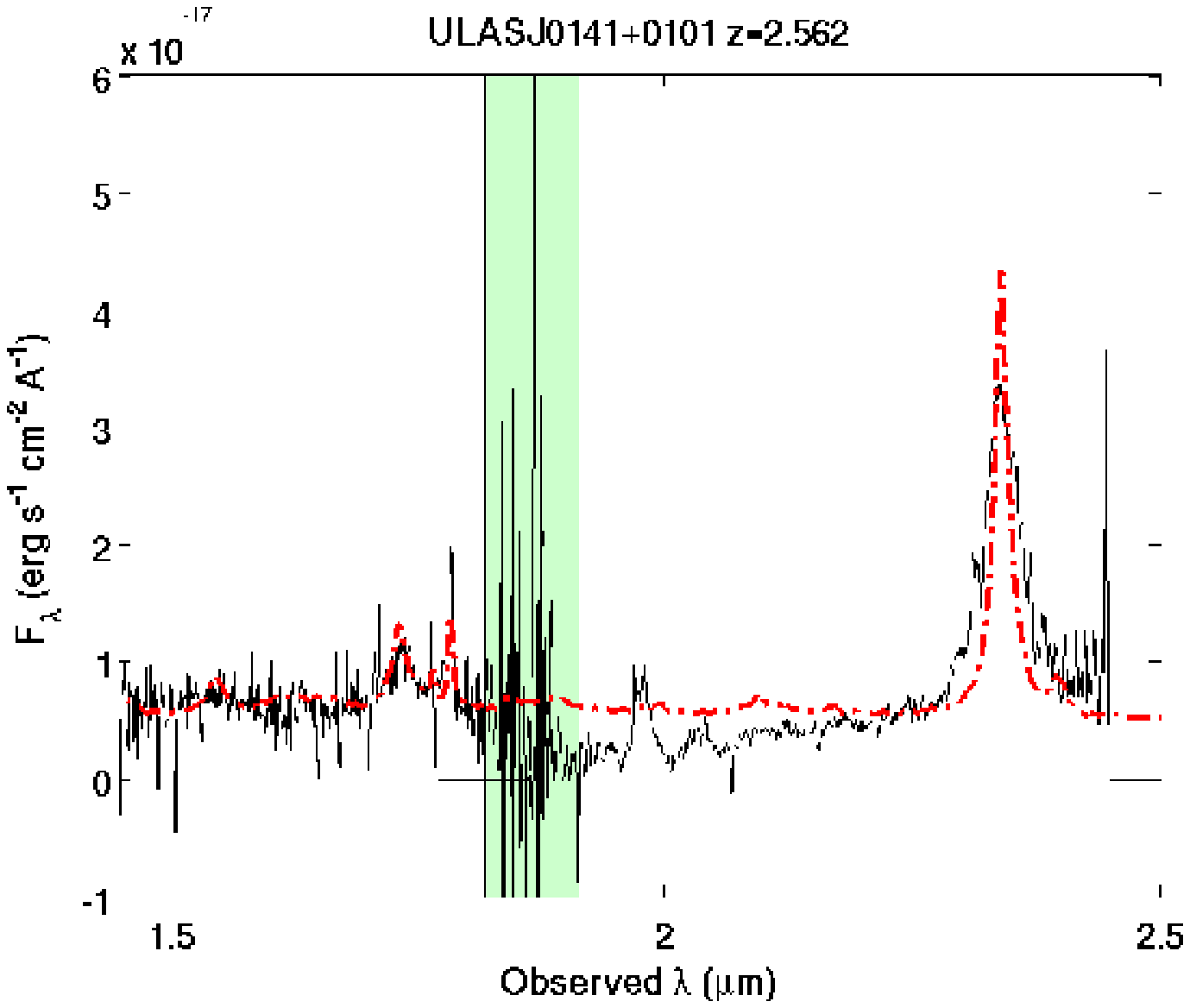} \\
\includegraphics[scale=0.35,angle=0]{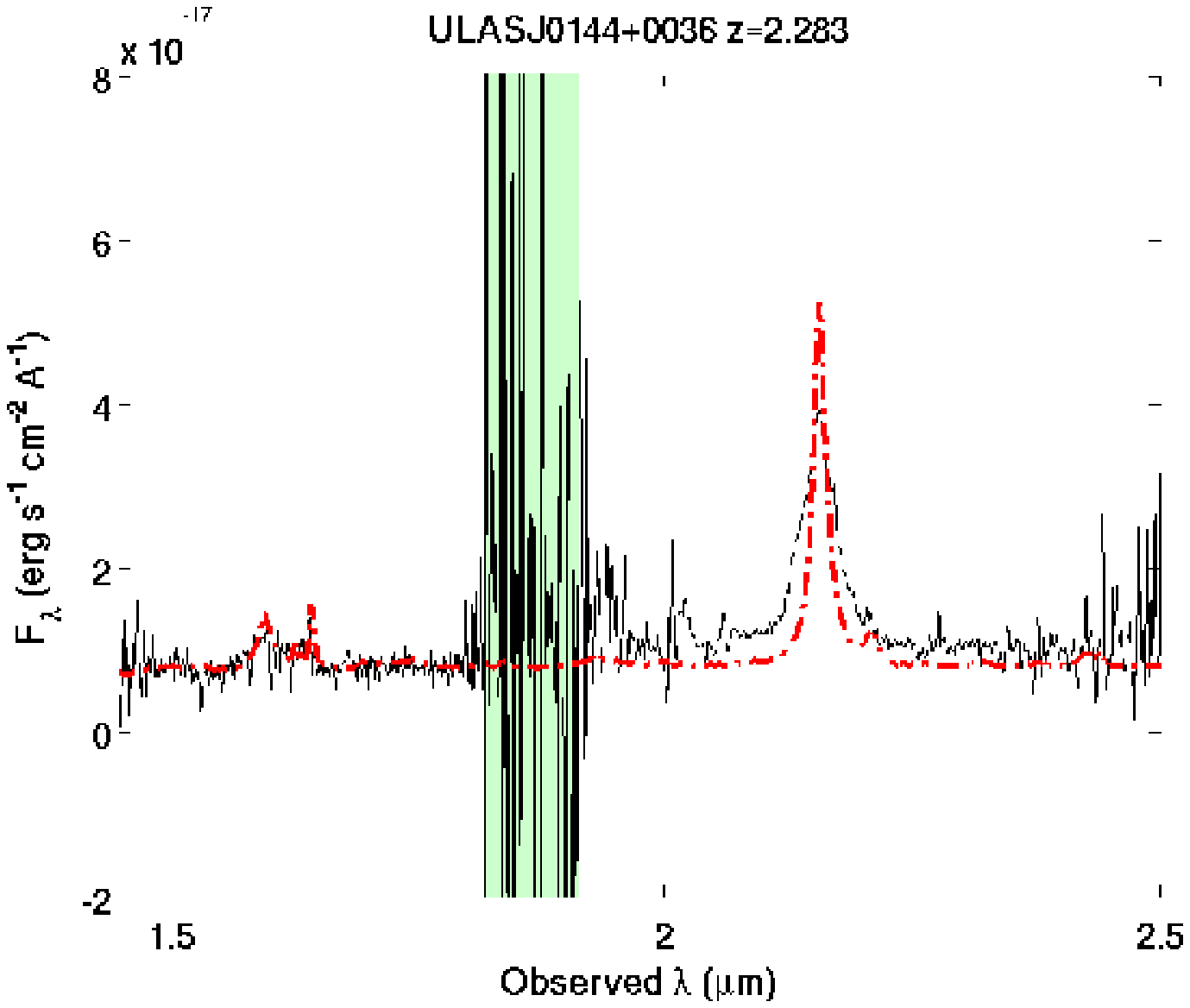} & \includegraphics[scale=0.35,angle=0]{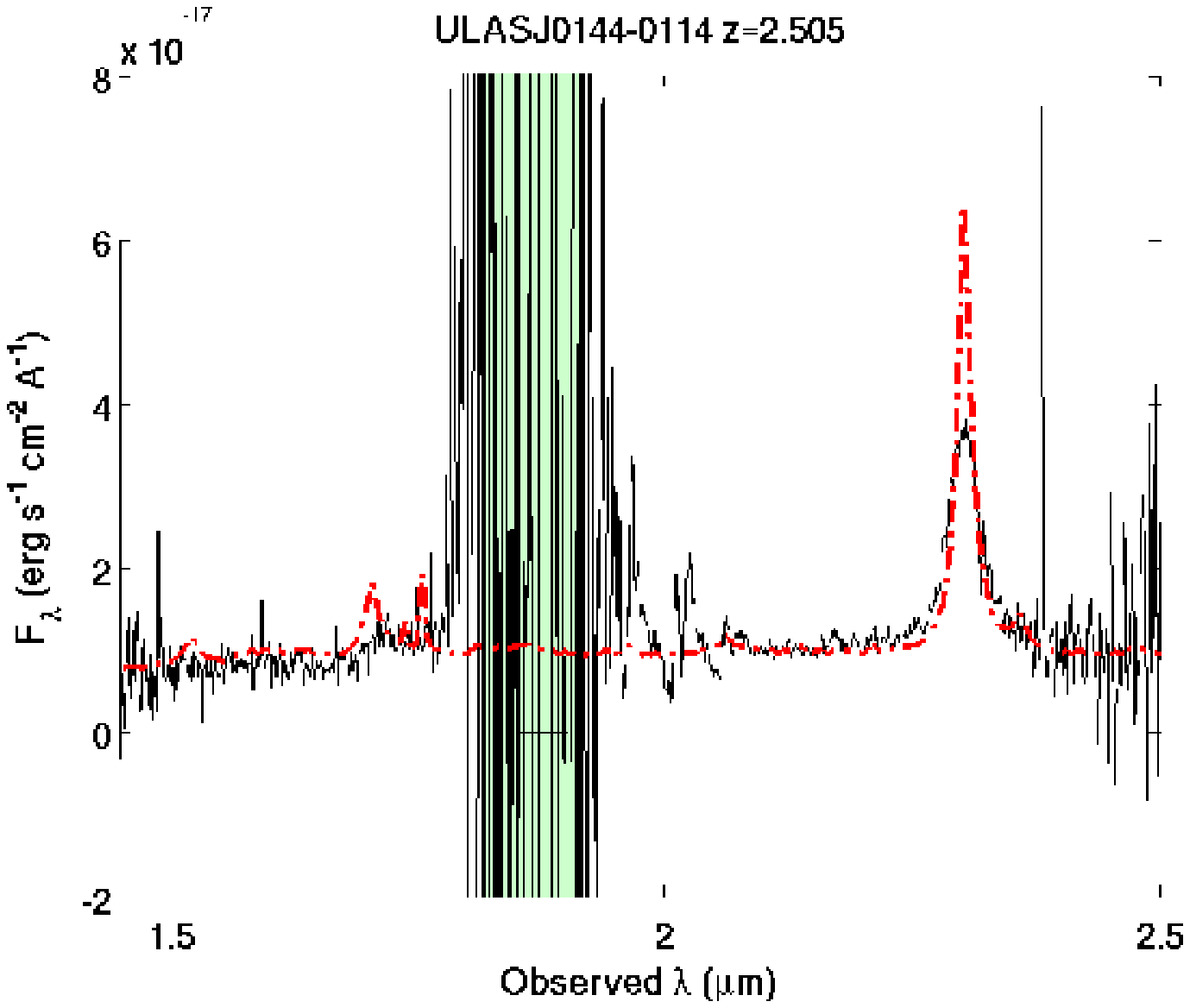} & \includegraphics[scale=0.35,angle=0]{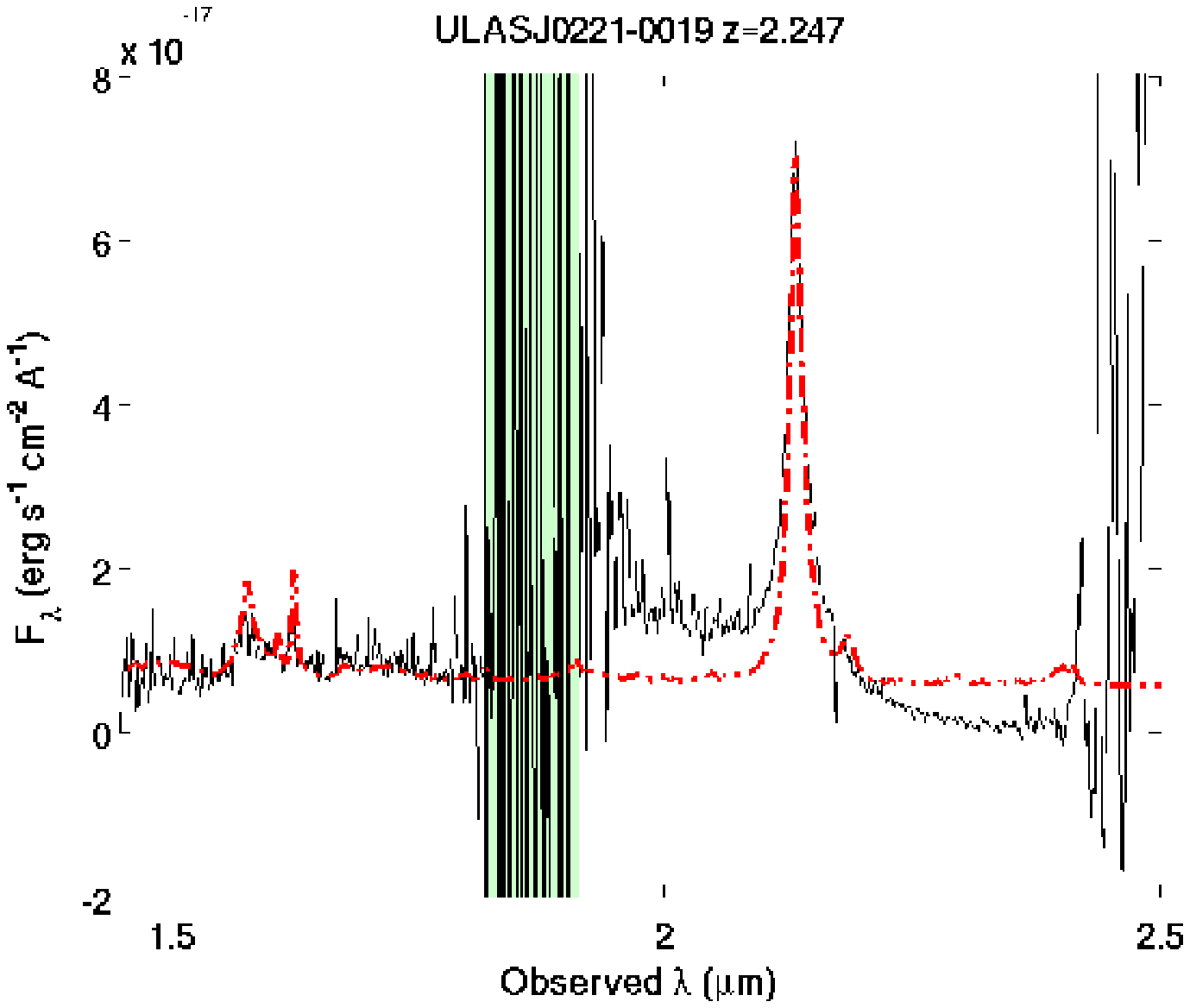} \\
\end{tabular}
%\includegraphics[width=9.5cm,height=7cm,angle=0]{./figs/sky_emission.eps}
%\end{tabular}
\end{minipage}
\caption{Near infra-red spectra of all 12 confirmed reddened quasars with redshifts
. The top two rows show VLT-SINFONI spectra apart from ULASJ1539+0557 which was observed using NIRSPEC on Keck. The bottom two rows show Gemini-GNIRS spectra. In the case of the GNIRS spectra, the $H$ and $K$-band orders are plotted together in these figures but problems with extracting the spectra at the edges of each order mean that the relative flux calibration between the two orders is unlikely to be accurate. All spectra have been smoothed by a 3--pixel boxcar
(0.0015\,$\mu$m) for purposes of presentation. The red dot-dashed lines
show the model fits derived for each quasar by fitting to the
broadband colours. The model SEDs are derived by fitting to the broadband colours of the quasars with a simple scaling of H$\alpha$ equivalent width to match the observations. The H$\alpha$ line profiles in the models and observations are therefore not expected to match. The region
of low atmospheric transmission between the $H$-- and $K$-bands is
shown as the grey rectangular area.}
\label{fig:spectra}
\end{center}
\end{figure*}

\begin{figure}
\begin{center}
\centering
\begin{minipage}[c]{1.00\textwidth}
%\begin{tabular}{c}
\includegraphics[width=9.5cm,height=7cm,angle=0]{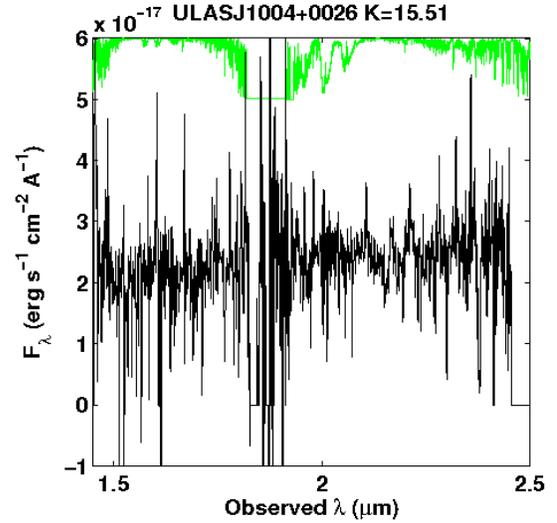}
%\end{tabular}
\end{minipage}
\caption{SINFONI spectrum of our brightest unconfirmed target
ULASJ1004+0026 which is unlikely to be an elliptical or low-redshift radio
galaxy. The atmospheric transmission spectrum from VLT-ISAAC is also
shown for reference at the top of the panel and demonstrates the wavelength region
(1.80--1.97$\mu m$ corresponding to H$\alpha$ at 1.75$<z<$2) where no lines can be
observed due to atmospheric opacity.}
\label{fig:nospectra}
\end{center}
\end{figure}

Three of the 13 quasar candidates observed show broad H$\alpha$ emission
in the $K$-band and two have H$\alpha$ emission in the $H$-band. We fit
multiple Gaussian components to the H$\alpha$ line profile and
redshifts are determined from the centroid of the intermediate
component where present and from the centroid of the single Gaussian where only one component is required to fit the data. Redshifts are summarised in Table
\ref{tab:sample}. In Fig.~\ref{fig:spectra} we show spectra of the
five new red quasars with redshifts as well as the seven confirmed quasars already presented in Paper I.

Out of the eight sources for which no redshifts were obtained, two
have associated FIRST counterparts so these could be lower redshift
radio galaxies with compact morphologies. One of the X-ray sources is
at the faint-end of our sample and below the nominal flux
limit. However, the other X-ray source, ULASJ1002+0137, which also lies
below the nominal flux limit of the survey was spectroscopically
confirmed to be a broad-line quasar at $z=1.595$. This source is in the COSMOS field and therefore has ancillary multi-wavelength data. We discuss it in detail in $\S$ \ref{sec:notes}.

In Sample 2, all the sources that had $i_{AB}>$20.5, were
spectroscopically confirmed to be reddened Type 1 AGN. No redshifts
were obtained for the four sources with $i_{AB}<=$20.5, consistent with
these being compact elliptical galaxies. We note that three out of the
four sources - ULASJ2148$-$0011, ULASJ2150$-$0055 and ULASJ2251+0047 - are indeed
classified as galaxies in the SDSS imaging survey. We use the publicly
available photometric redshift code \textsc{Eazy} \citep{Brammer:08}
to calculate photometric redshifts for these four SDSS-bright EROs
using an $R$-band luminosity function prior. The photometric redshifts
of the four sources are summarised in Table \ref{tab:photoz}.

%In Figure \ref{fig:2150} we show the best-fit SED for ULASJ2150 for the flat-prior solution of z=4.77 along with the photometric datapoints and the observed spectrum. 

%\begin{figure}
%\begin{center}
%\includegraphics[scale=0.5,angle=0]{./figs/ULASJ2150_zp462.eps}
%\caption{Best-fit SED to ULASJ2150. No emission lines are seen in the infra-red spectrum and the broadband photometry is consistent with this source being a compact elliptical galaxy at z=4.77.}
%\label{fig:2150}
%\end{center}
%\end{figure}

The sources without redshifts which are also undetected in the optical
filters could have several interpretations. The spectrum for the
brightest of these, ULASJ1004+0026, is shown in Fig.~\ref{fig:nospectra}
along with the sky transmission spectrum. If such unconfirmed candidates
are quasars, as suggested by their compact
morphologies and NIR colours, they may be at redshifts between
1.75--2.00 which corresponds to H$\alpha$ lying in the region of strong atmospheric transmission between the $H$ and $K$ bands. Alternatively, there is a class of rare broad absorption
line quasars known as FeLOBALs \citep{Voit:93, Becker:97, Becker:00, Hall:02} where strong absorption
due to iron at relatively low ionisation potentials means there is
virtually no flux below $\lambda \sim 2800$\,\AA\@. It has been
hypothesised that FeLOBALs may represent a transition population
between ultra-luminous infra-red galaxies and quasars with some of
these sources having prodigiously high star formation rates, as
inferred from their mid infra-red properties
\citep{Farrah:07}. Although they are very rare, they would be
preferentially selected in NIR-selected samples due to their very red
colours - e.g. as in \citet{Urrutia:09}. If some of our unidentified objects do indeed correspond to
this class of objects, this would place them at $z\sim$3--4 in order
to reproduce the very red colours. This is consistent with the fraction of BALs increasing at these higher redshifts \citep{Allen:11}. Optical spectroscopy would help
confirm the identity of these sources.

%Finally we point out that ULASJ2150, one of our brightest non-identifications does seem to have absorption features in the near infra-red bands seen in Figure \ref{fig:nospectra}. If these features are real, this source may be a very compact high redshift elliptical galaxy with a 4000\AA\@ break and Balmer absorption lines.  

\begin{table}
\begin{center}
\caption{Photometric redshifts for the four SDSS Stripe82 EROs with
$i<=$20.5 and for which no lines were identified in the NIR
spectra.}
\label{tab:photoz}
\begin{tabular}{lc}
\hline \hline
Name & $z_{\rm{p}}$\\
\hline
ULASJ2148$-$0011 & 0.97$\pm^{0.08}_{0.09}$ \\
ULASJ2150$-$0055 & 1.30$\pm$0.08 \\
ULASJ2219+0036 & 2.71$\pm$0.06 \\
ULASJ2251+0047 & 0.85$\pm$0.05 \\
\hline
\end{tabular}
\end{center}
\end{table}

\section{RESULTS}

\label{sec:results}

The spectra presented above combined with the sample in Paper I total
12 confirmed Type 1 AGN with very red colours and redshifts between
1.4--2.7. We now turn to a detailed assessment of their properties.

\subsection{Model SEDs}

\label{sec:sed}

We begin by fitting model Type 1 AGN SEDs to the available SDSS+UKIDSS broadband photometry for all our quasars. The photometric data is chosen for the fitting as it covers a larger wavelength range than our near infra-red spectra and includes flux points at bluer wavelengths which are more sensitive to dust extinction. The
unreddened 'standard' quasar model is described in Section~2.4 of
\citet{Maddox:08}.  The model SED reproduces the SDSS $ugriz$ and
UKIDSS $YJHK$ broadband colours of the majority of quasars, selected
in the SDSS, to high accuracy over the redshift range $0.2 \le z \le
5$. The aim of fitting these model SEDs to the quasar photometry, is to infer the amount of dust extinction in each quasar from the observed colours. Although other factors such as lensing and variability may also contribute to the red colours, these are unlikely to be significant in our sample. All our quasars lie at z$\gtrsim$1.5 and are very luminous so the host galaxy contribution to the broadband light is likely to be insignificant. While variability can certainly affect the spectral shape of quasars, the effect is more pronounced in the rest-frame UV bands and lower luminosity objects \citep{Hook:94} and more often seen in radio-loud populations \citep{Helfand:01}. Only two of our 12 quasars, ULASJ0016$-$0038 and ULASJ0141+0101 are radio-loud. We discuss the possibility of lensing in detail in $\S$ \ref{sec:lens} where we conclude that it is unlikely to affect the majority of our sample. We therefore proceed by assuming that the red NIR colours seen in our quasars is due to dust extinction.

Reddened model SEDs were therefore created for each quasar, by including the effect of an
extinction curve with a specified $E(B-V)$. The form of the extinction
curve used is very similar to that of \citet{Gallerani:10} but in the
rest-frame wavelength interval of interest here, 2500-10\,000\,\AA,
the differences between commonly used extinction curves (e.g. LMC, SMC
and Milky Way) are small and the results are not dependent on the
choice of extinction curve. The facility to adjust the emission line
equivalent widths is also incorporated in the models and
object-specific model SEDs were generated by matching the measured
H$\alpha$ equivalent widths. We stress however, that we do not attempt to match the H$\alpha$ line profiles in the model and observed spectra apart from this simple scaling of the equivalent widths. In this way, we can disentangle the effects of H$\alpha$ equivalent width and dust extinction in the observed $(J-K)$ colours. 

In a few cases, the $(J-K)$ and $(H-K)$ colours are marginally
inconsistent with a single reddening value so we quote the best-fit
extinction values needed to reproduce each of the observed
colours. However, the inferred $E(B-V)$ values are always similar and
the conclusions of the paper do not depend critically on which values
are adopted. For the rest of the paper, we use the extinction values 
derived by matching to the $(J-K)$ colours so as to provide a larger 
baseline in terms of wavelength for the extinction estimates. 
An R$_V$=3.1, consistent with the dust in our own Milky
Way, is assumed throughout\footnote{Considering possible variations in the
dust-law in distant quasars is beyond the scope of this work.}.

In Table \ref{tab:modelfit} we quote the dust extinction, $A_V$ inferred from the model fits for
our combined sample of reddened quasars from Paper I and the current
work. We also give the factor by which the H$\alpha$ equivalent widths
in the model are scaled in order to reproduce the
observations. Previous studies by \citet{Glikman:07} did not account
for the effect of the H$\alpha$ equivalent widths on the $(J-K)$
colours. Fig.~\ref{fig:spectra} shows these best-fit model SEDs
overlaid on the observed rest-frame optical spectra for all our spectroscopically confirmed red quasars. Note that the model fits were derived from the broadband photometry only with a simple scaling adopted to match the observed H$\alpha$ equivalent widths so the match between these model SEDs and the observed continuum flux in the spectra is, in most cases, very good. In some cases - e.g. ULASJ0221$-$0019 - the models do not match the continuum in the observed spectra at the edges of the spectral orders due to problems with extracting the spectra at the edges. We see from Table \ref{tab:modelfit} however that the match between the observed and model colours is always very good. The median difference in the model and observed $(H-K)$ colours is 0.005.

\begin{table*}
\begin{center}
\caption{Summary of Model Fits for Spectroscopically Confirmed Red
Quasars.}
\label{tab:modelfit}
\begin{tabular}{lccccccc}
\hline \hline
Name & Redshift & $(J-K)_{\rm{obs}}$ & $(J-K)_{\rm{model}}$ & $(H-K)_{\rm{obs}}$ & $(H-K)_{\rm{model}}$ & A$_V$ & H$\alpha$ EW Scaling \\
\hline
ULASJ0016$-$0038 & 2.194 & 1.607 & 1.702 & 0.871 & 0.882 & 1.7 & 2.1 \\
ULASJ0041$-$0021 & 2.517 & 2.171 & 1.821 & 0.851 & 0.869 & 2.1 & 0.7 \\
          & 2.517 & 2.171 & 2.167 & 0.851 & 1.023 & 2.5 & 0.7 \\
ULASJ0141+0101 & 2.562 & 1.767 & 1.752 & 0.893 & 0.893 & 1.8 & 1.2 \\
ULASJ0144$-$0114 & 2.505 & 1.852 & 1.852 & 1.009 & 0.897 & 2.1 & 1.0 \\
               & 2.505 & 1.852 & 2.099 & 1.009 & 1.007 & 2.4 & 1.0 \\
ULASJ0144+0036 & 2.283 & 1.902 & 1.496 & 0.726 & 0.734 & 1.7 & 0.9 \\
               & 2.283 & 1.902 & 1.901 & 0.726 & 0.913 & 2.3 & 0.9 \\
ULASJ0221$-$0019 & 2.247 &  -- & 1.690 & 0.854 & 0.856 & 1.8 & 1.6 \\
ULASJ1539+0557 & 2.658 & -- & 3.290 & 1.440 & 1.450 & 4.0 & 0.3 \\
ULASJ1002+0137 & 1.595 & -- & 1.657 & 0.551 & 0.550 & 3.2 & 0.8 \\
ULASJ1234+0907 & 2.503 & -- & 5.065 & 2.321 & 2.331 & 6.0 & 1.3 \\
ULASJ1455+1230 & 1.460 & 2.212 & 1.729 & 0.570 & 0.571 & 3.4 & 0.9 \\
          & 1.460 & 2.212 & 2.219 & 0.570 & 0.790 & 4.3 & 0.9 \\
ULASJ2200+0056 & 2.541 & 1.693 & 1.617 & 0.798 & 0.815 & 1.7 & 1.3 \\
ULASJ2224$-$0015 & 2.223 & 1.747 & 1.798 & 0.910 & 0.912 & 1.9 & 1.7 \\
\hline
\hline
\end{tabular}
\end{center}
\end{table*}

From Table \ref{tab:modelfit} it can immediately be seen that most of
the red quasars have similar H$\alpha$ equivalent widths to the
standard model spectrum constructed from the optically selected quasar
population, with the strongest H$\alpha$ line having an equivalent
width of only a factor of two larger. However, we find that significant
dust extinction is required to reproduce the continuum colours, with a
mean dust extinction of A$_V \sim$2.5 for all our sources. ULASJ1234+0907,
our reddest object, has a dust extinction of A$_V$=6.0 inferred from its ($H-K$) colour and is one of
the reddest quasars currently known.

\subsection{Line Widths, Bolometric Luminosities \& Black Hole Masses}

\label{sec:bh}

\begin{figure*}
\begin{center}
\centering
\begin{tabular}{cc}
\includegraphics[width=8.5cm,height=6.0cm,angle=0]{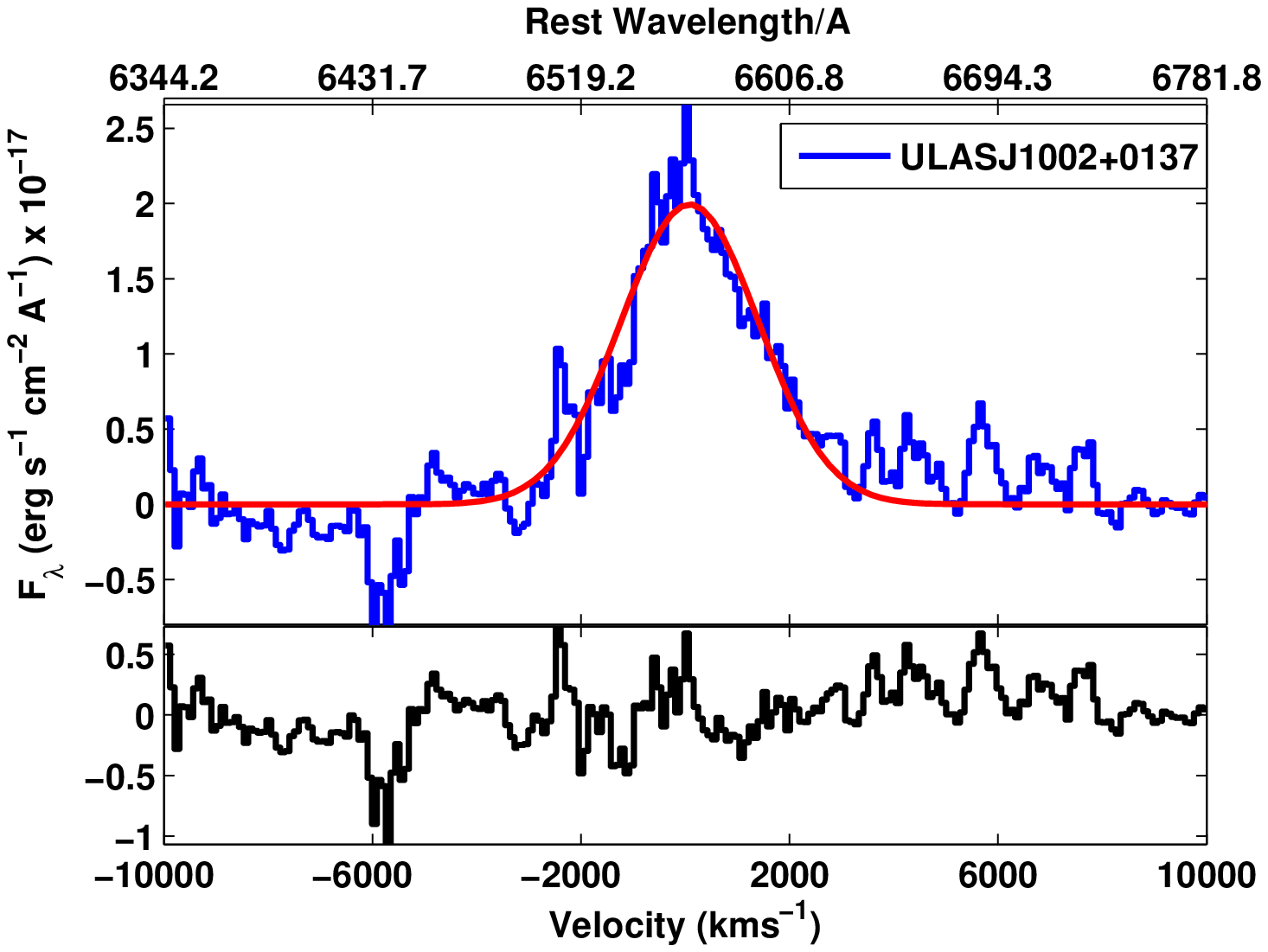} & \includegraphics[width=8.5cm,height=6.0cm,angle=0]{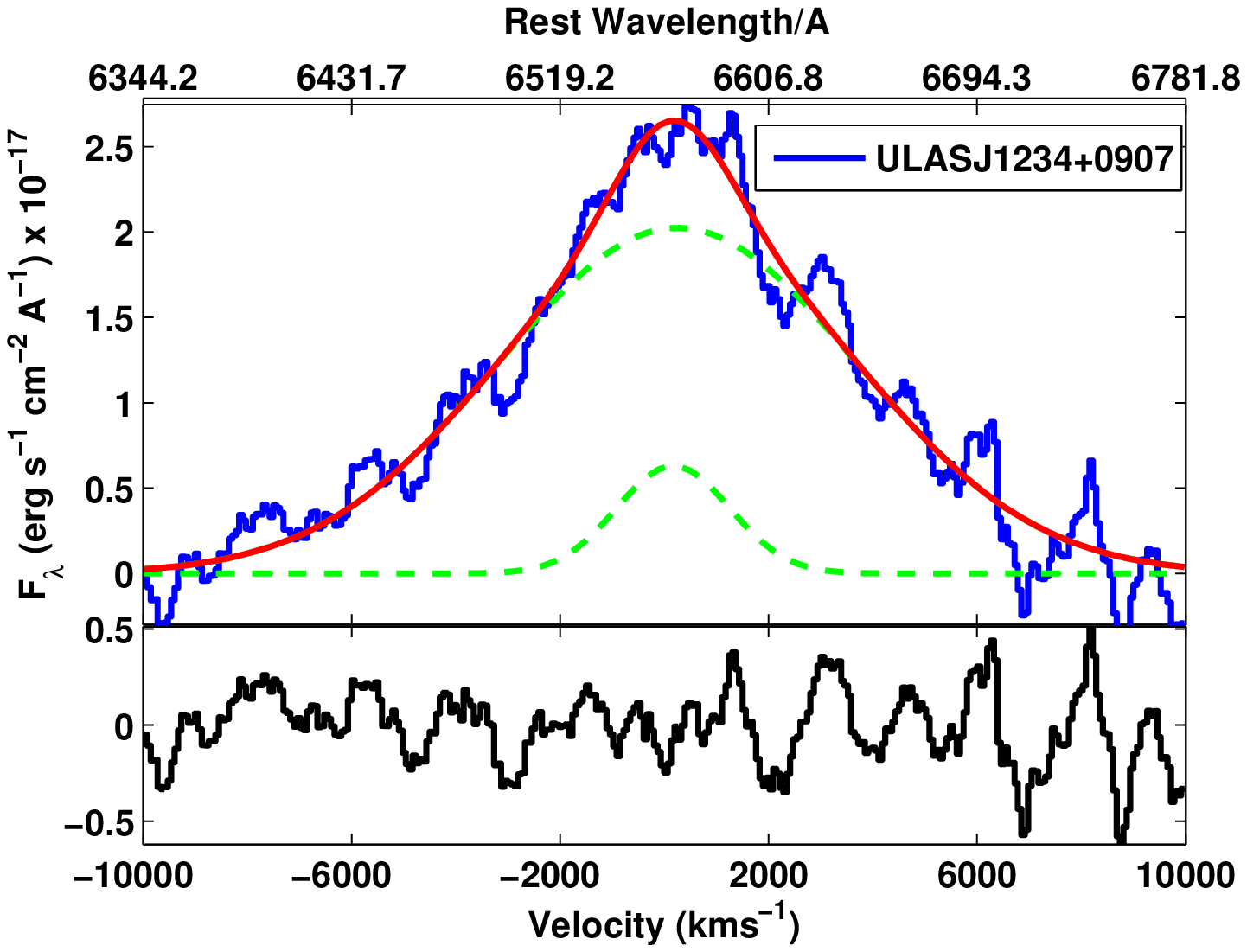} \\
\includegraphics[width=8.5cm,height=6.0cm,angle=0]{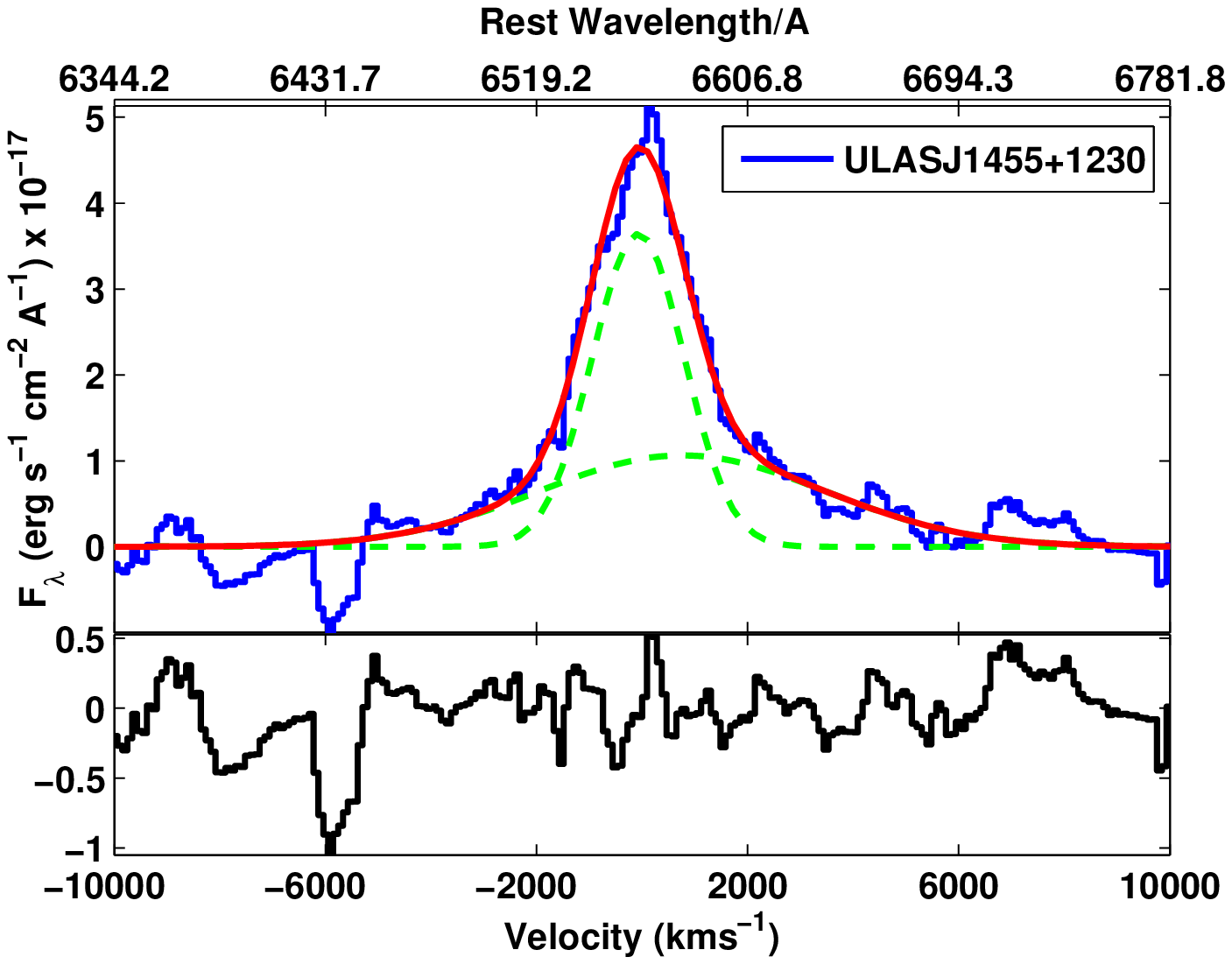} & \includegraphics[width=8.5cm,height=6.0cm,angle=0]{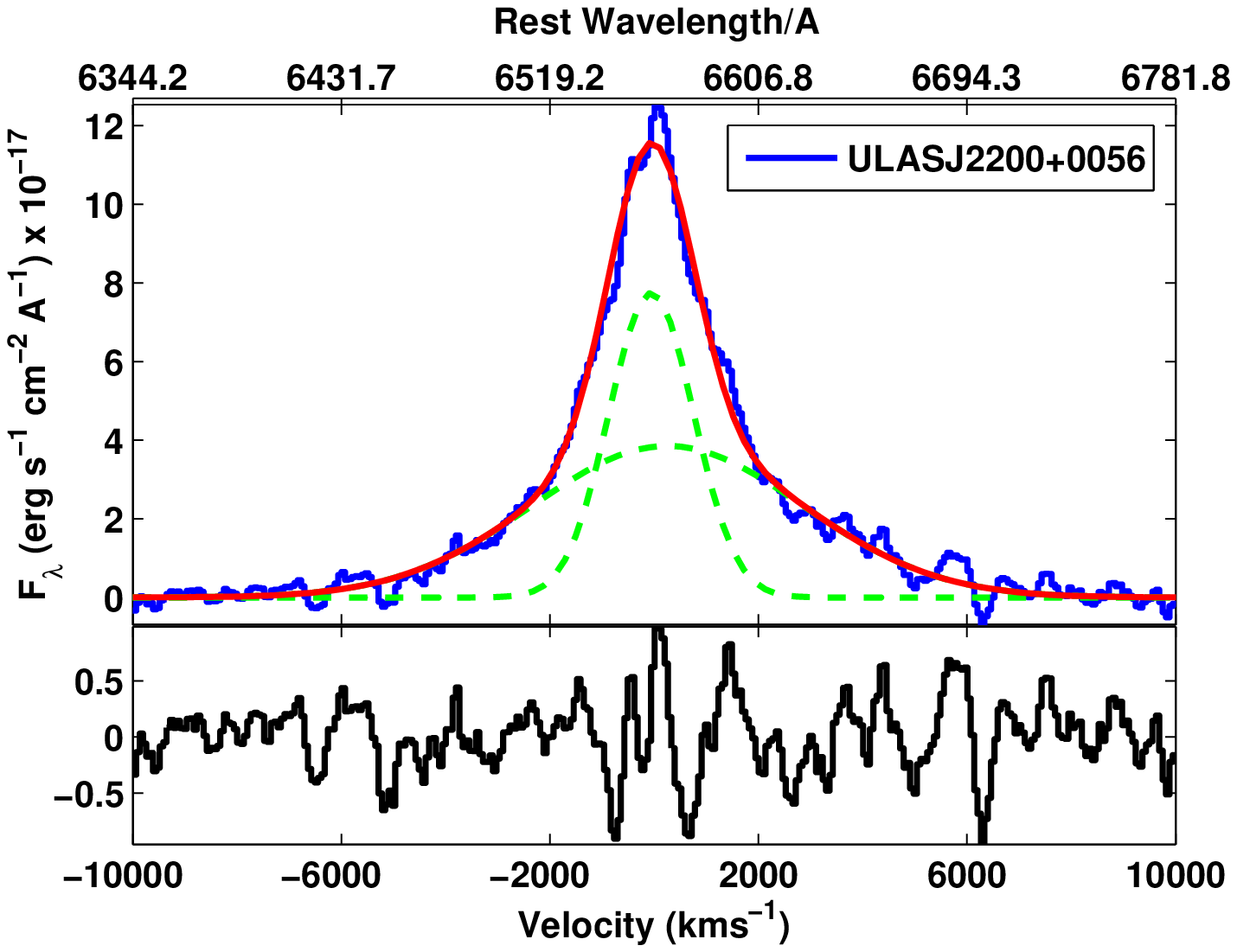} \\
\end{tabular}
\begin{tabular}{c}
\includegraphics[width=8.5cm,height=6.0cm,angle=0]{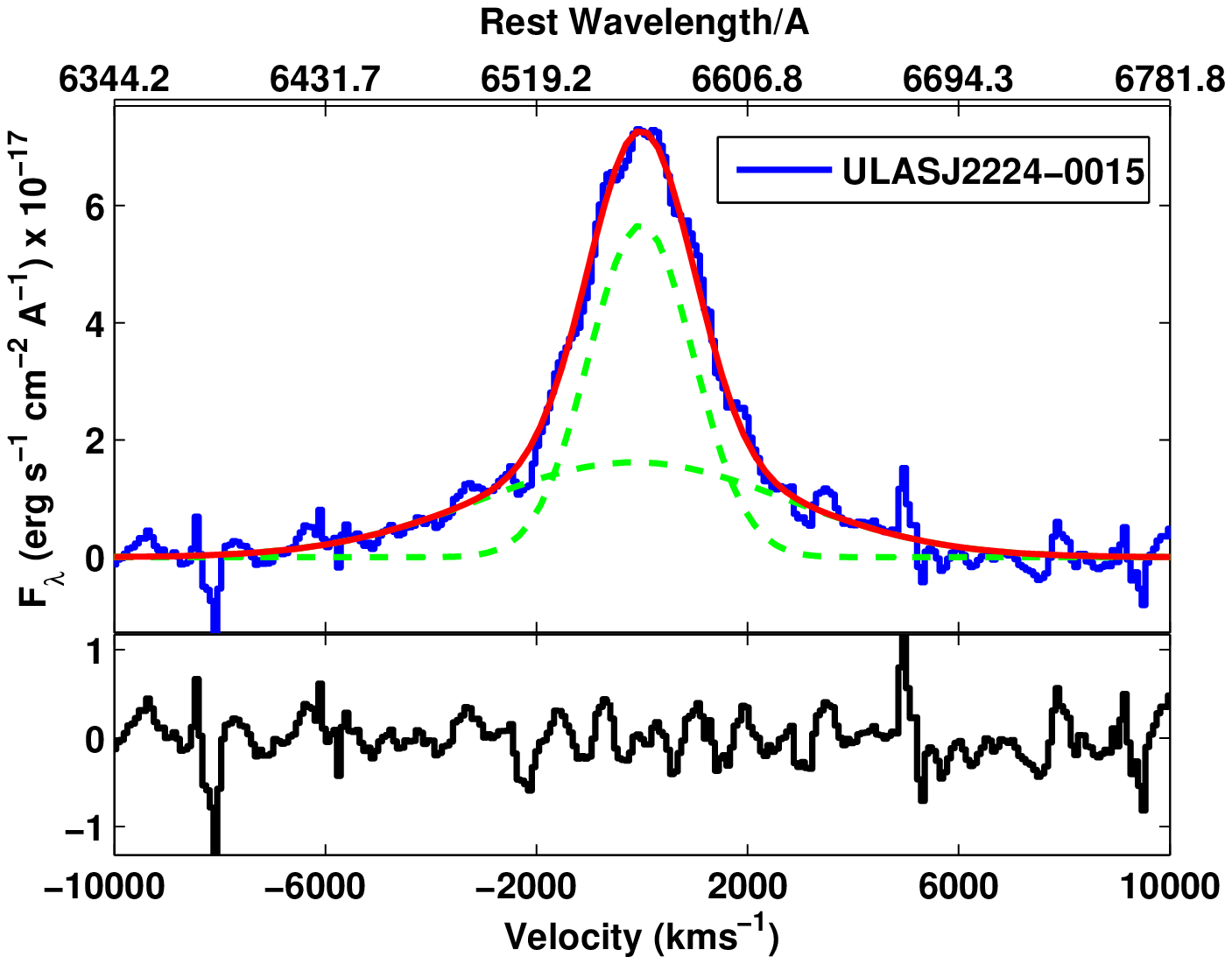} \\
\end{tabular}
\caption{Gaussian fits to H$\alpha$ line for the five new quasars. All
observed spectra have been smoothed by a 5-pixel boxcar for purposes
of presentation. A broad as well as an intermediate component are
needed to fit the line profiles adequately for most of the
sources. These individual components are shown as the dashed lines
whereas the solid lines show the sum of the two components. We note
that in the case of ULASJ1002+0137 a single-component fit, with
FWHM=3200$\pm$300kms$^{-1}$, produced an equally good $\chi^2$. The
residuals from the fits are also shown at the bottom of each panel.}
\label{fig:Hagauss2}
\end{center}
\end{figure*}

\begin{figure*}
\begin{center}
\centering
\begin{tabular}{cc}
\includegraphics[width=8.5cm,height=6.0cm,angle=0]{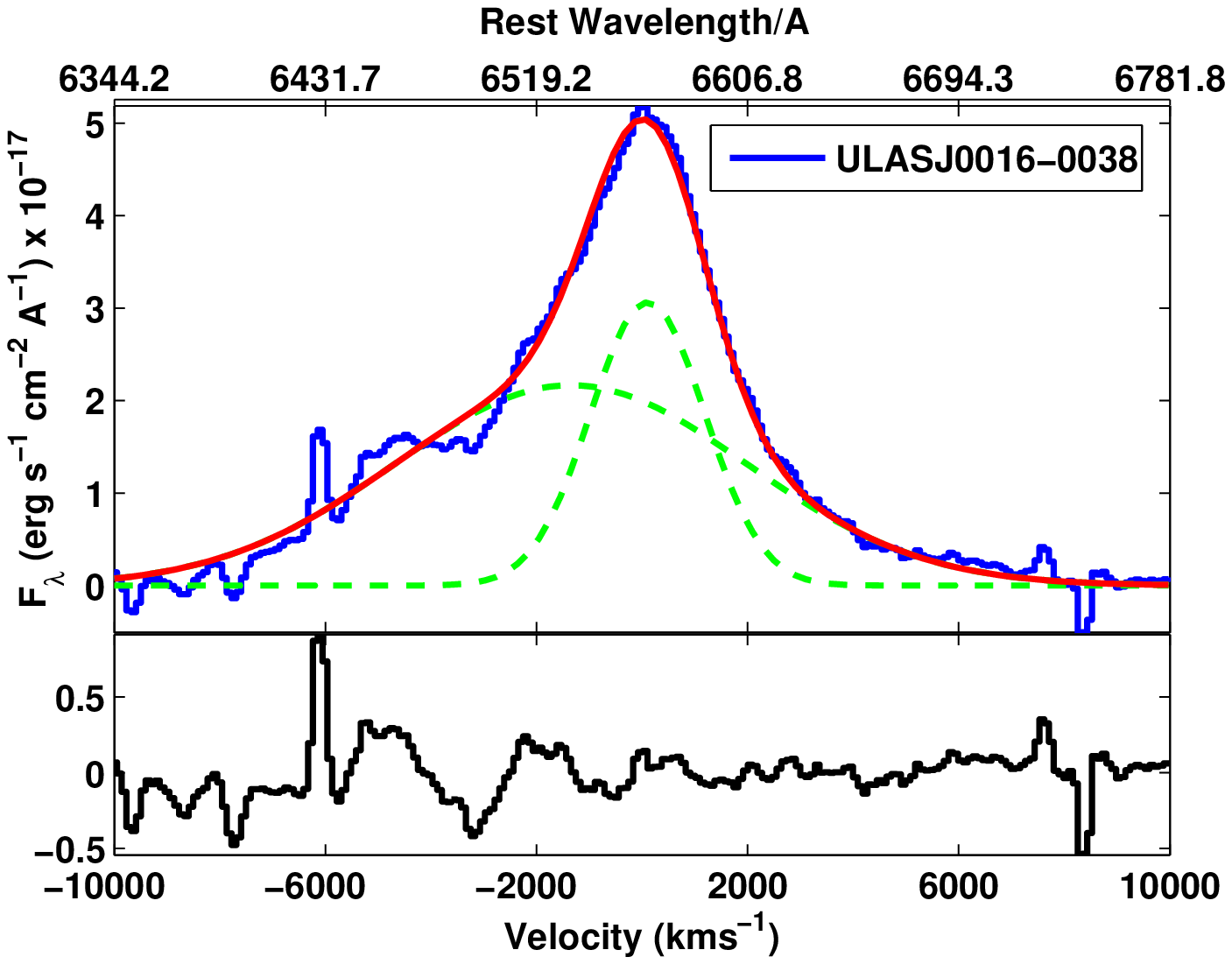} & \includegraphics[width=8.5cm,height=6.0cm,angle=0]{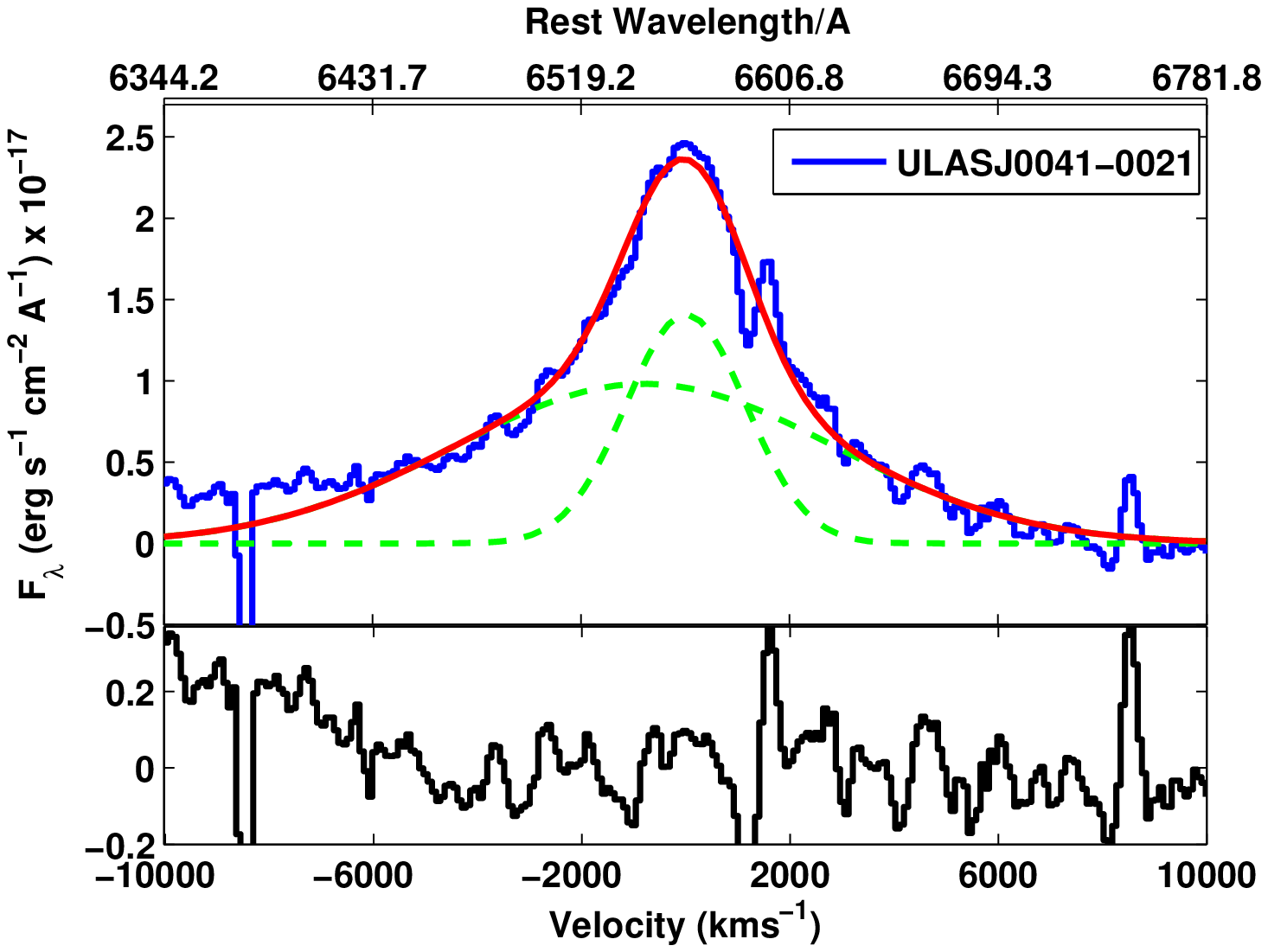} \\
\includegraphics[width=8.5cm,height=6.0cm,angle=0]{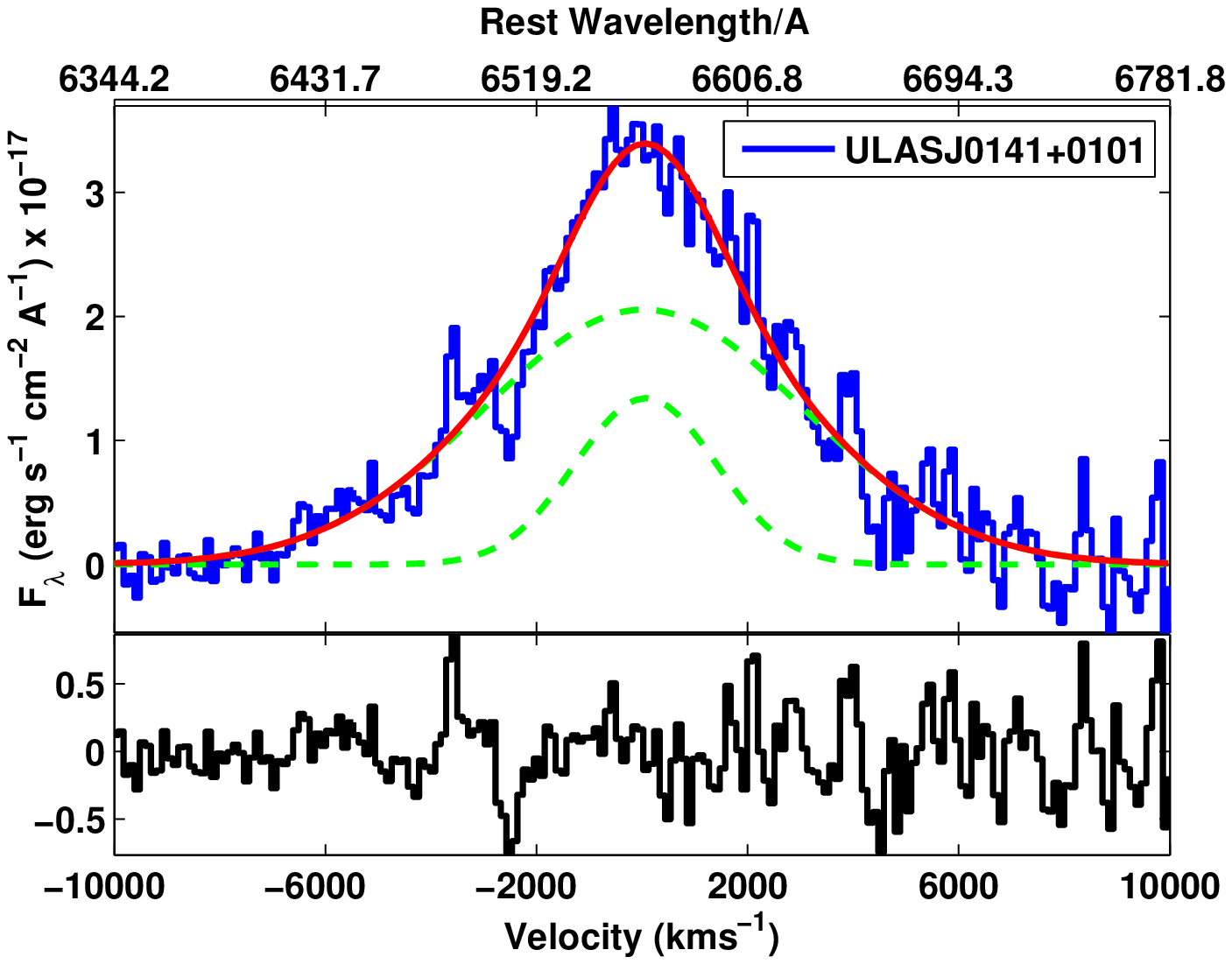} & \includegraphics[width=8.5cm,height=6.0cm,angle=0]{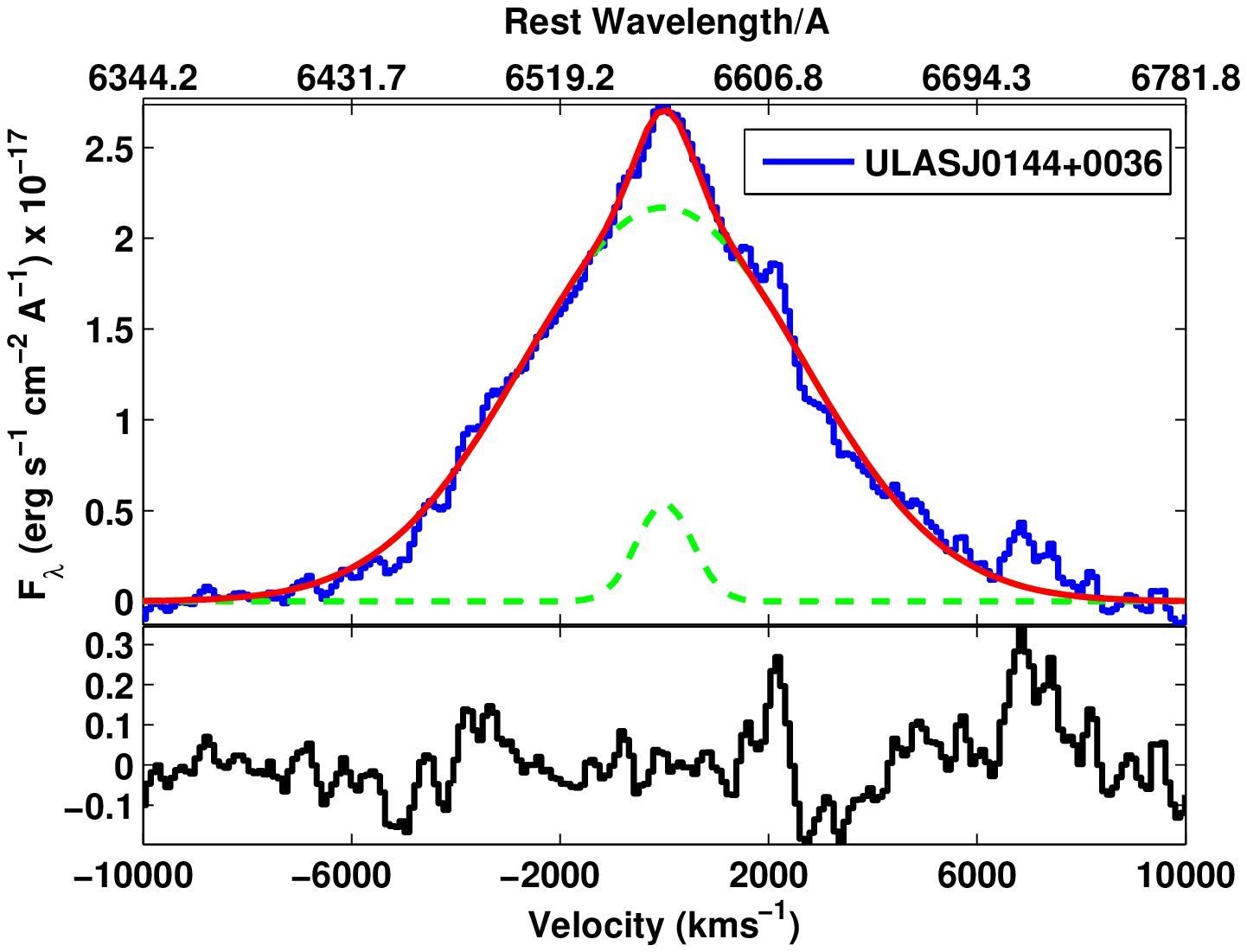} \\
\includegraphics[width=8.5cm,height=6.0cm,angle=0]{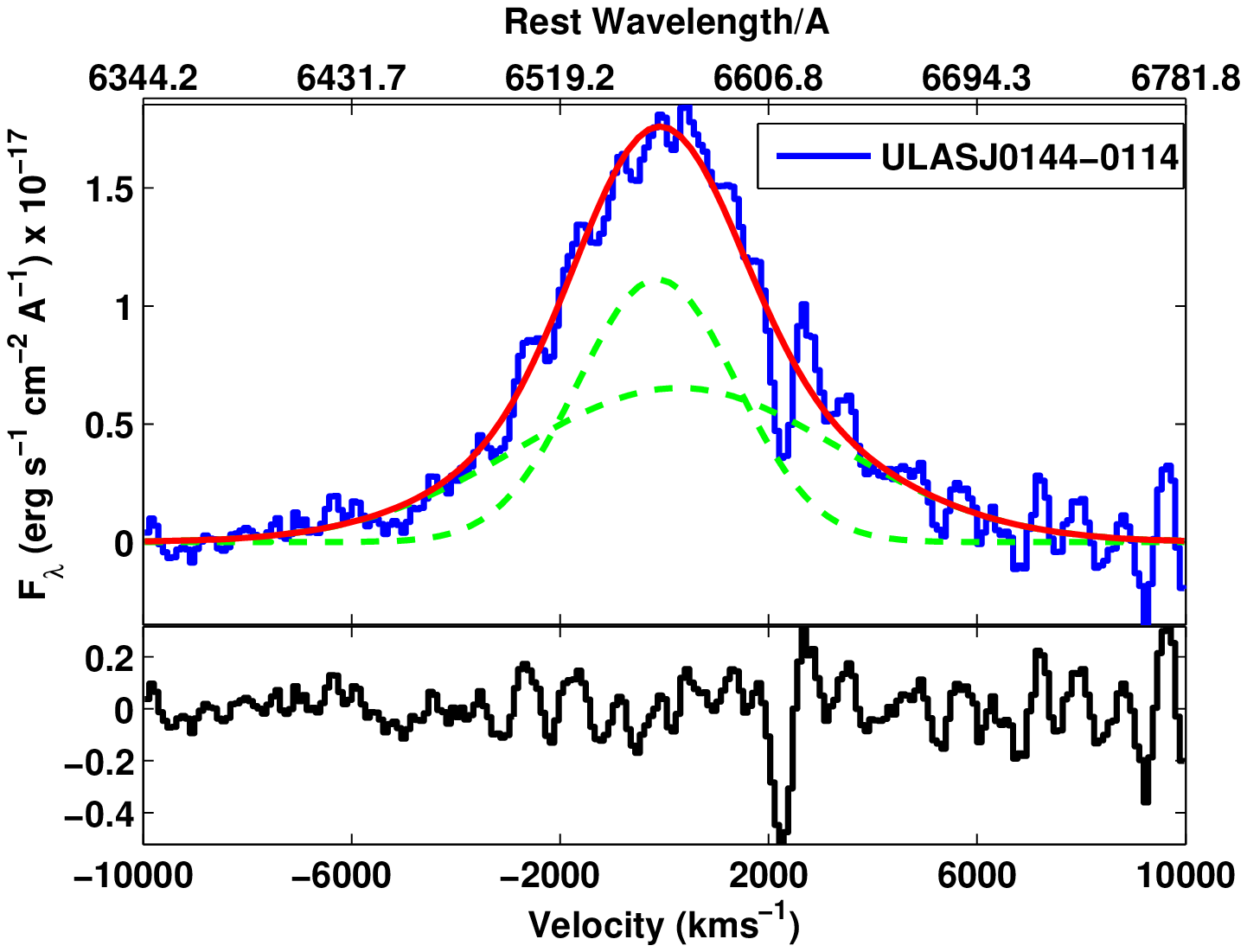} & \includegraphics[width=8.5cm,height=6.0cm,angle=0]{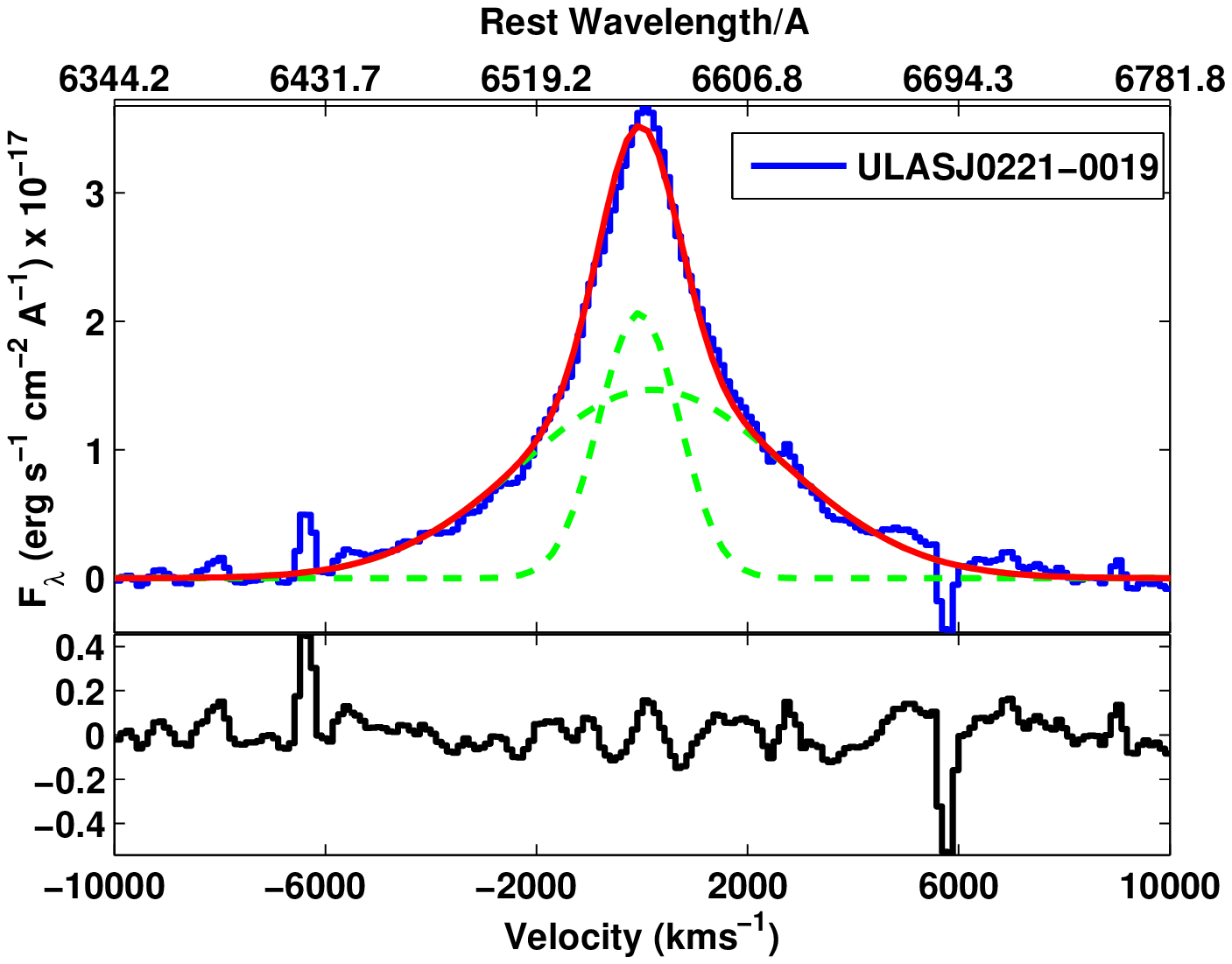}
\end{tabular}
\begin{tabular}{c}
\includegraphics[width=8.5cm,height=6.0cm,angle=0]{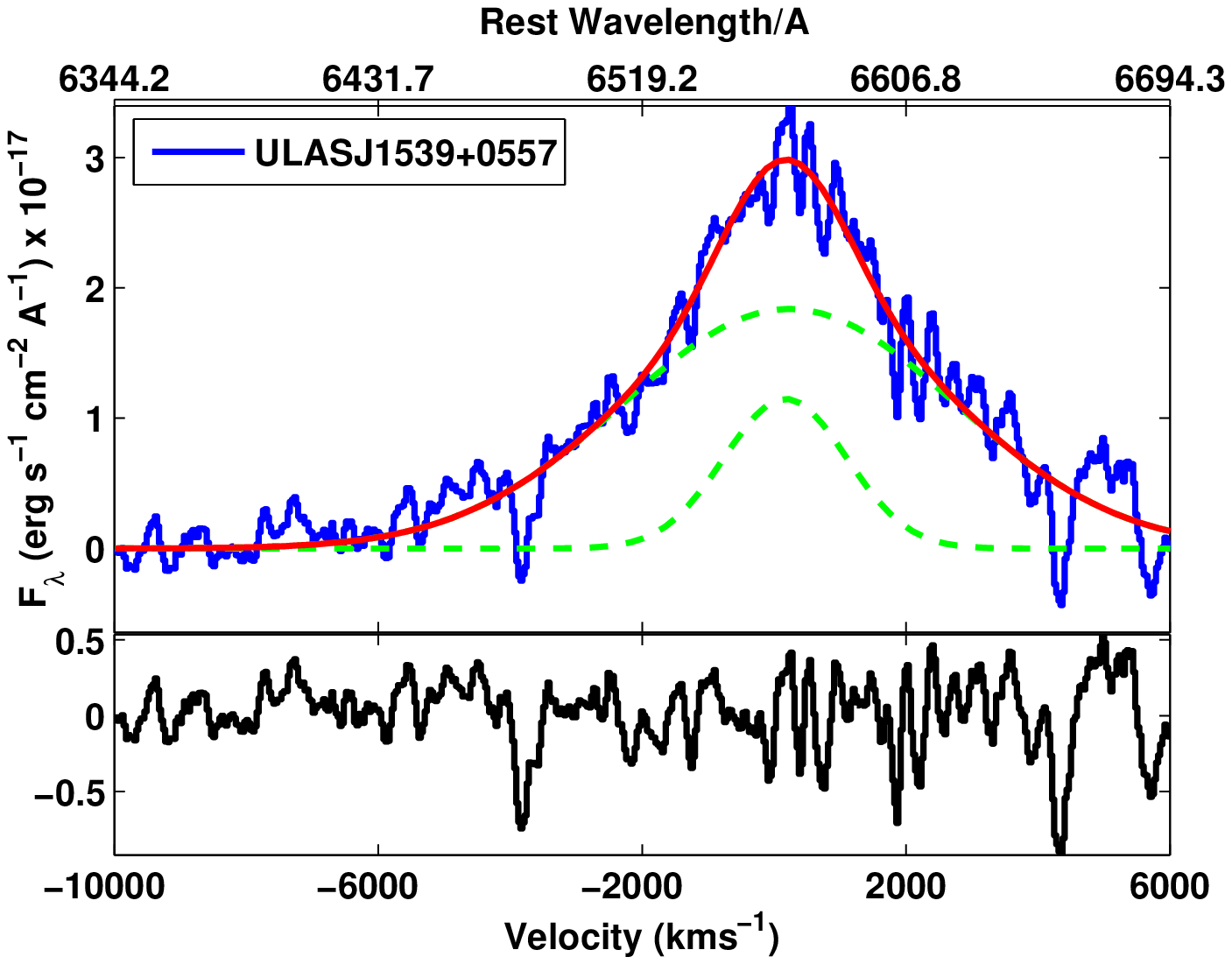} \\
\end{tabular}
\caption{As in Figure \ref{fig:Hagauss2} but for the Paper I sample.}
\label{fig:Hagauss}
\end{center}
\end{figure*}

In this section we assess the emission line properties of our 12 red
quasars. In order to do so, we fit multiple Gaussian components to the
H$\alpha$ line profiles of all our quasars with spectroscopic
redshifts after blueshifting to the rest-frame. Note, these Gaussian fits to the observed H$\alpha$ lines are distinct from the model SED fits done to the broadband photometry in $\S$ \ref{sec:sed} to infer a dust extinction. The effect of this dust extinction on the full width half maximum (FWHM) of the H$\alpha$ line is expected to be very small. The Gaussian fits can
be seen in Fig.~\ref{fig:Hagauss2} and Fig.~\ref{fig:Hagauss} for the
new sample presented in this paper and the Paper I sample
respectively. The derived line-widths are summarised in Table
\ref{tab:Halpha}. Errors are derived from a least-squares fit of the
Gaussian profiles to the data. In most cases, both intermediate and
broad components are needed to fit the line profiles adequately. The
presence of a very broad H$\alpha$ component with FWHM $>$5000\,kms$^{-1}$ in these quasars, confirms that we
are able to at least partially see through into the broad line region
located close to the black-hole accretion disk. The broad components
in our sample typically have FWHM=6000\,kms$^{-1}$ while the
intermediate component has an average FWHM=2230\,kms$^{-1}$. Our red
quasars are therefore likely to be canonical Type 1 AGN rather than
Type 2 AGN that are obscured by a dusty torus
(e.g. \citet{Martinez-Sansigre:05}). In the latter case we would not
be able to detect the broad component for most lines of sight. We
conclude that the large dust extinction values must be arising in the
host galaxy of the quasar rather than from the molecular torus.

\begin{table*}
\begin{center}
\caption{H$\alpha$ FWHM derived by fitting both a
single Gaussian as well as two (broad and an intermediate components)
Gaussian to the H$\alpha$ line profiles. A narrow H$\alpha$ component 
was not required in any of the sources.}
\label{tab:Halpha}
\begin{tabular}{lccc}
\hline \hline
Name & FWHM$_{H\alpha}^{\rm{single}}$/kms$^{-1}$ & FWHM$_{H\alpha}^{\rm{broad}}$/kms$^{-1}$ & FWHM$_{H\alpha}^{\rm{int}}$/kms$^{-1}$ \\
\hline
ULASJ0016$-$0038 & 5400$\pm$400 & 7900$\pm$200 & 2620$\pm$70 \\
ULASJ0041$-$0021 & 5500$\pm$500 & 8700$\pm$600 & 2800$\pm$200 \\
ULASJ0141+0101 & 5900$\pm$300 & 7300$\pm$300 & 3200$\pm$100 \\
ULASJ0144+0036 & 6000$\pm$200 & 6400$\pm$200 & 1410$\pm$40 \\
ULASJ0144$-$0114 & 4700$\pm$400 & 5400$\pm$500 & 2800$\pm$200 \\
ULASJ0221$-$0019 & 3700$\pm$200 & 6000$\pm$100 & 1950$\pm$30 \\
ULASJ1539+0557 & 4700$\pm$400 & 5900$\pm$400 & 2200$\pm$200 \\
ULASJ1002+0137 & 3200$\pm$300 & -- & -- \\
ULASJ1234+0907 & 7300$\pm$600 & 8100$\pm$600 & 2600$\pm$200 \\
ULASJ1455+1230 & 2900$\pm$200 & 6400$\pm$300 & 2000$\pm$100 \\
ULASJ2200+0056 & 3300$\pm$200 & 6200$\pm$100 & 1960$\pm$40 \\
ULASJ2224$-$0015 & 3100$\pm$100 & 6700$\pm$200 & 2350$\pm$60 \\
\hline
\end{tabular}
\end{center}
\end{table*}

The H$\alpha$ line properties can be used to infer black-hole masses,
bolometric luminosities as well as Eddington fractions using the
scaling relations between H$\alpha$ and H$\beta$ linewidth
\citep{Greene:05} from reverberation mapping \citep{Kaspi:07}. We
correct the line for the instrumental broadening of 185\,kms$^{-1}$ by subtracting this in quadrature from the observed linewidth and use the equations below:

\begin{equation}
\rm{FWHM}_{H\beta}=(1.07\pm0.07)\times10^{3}\left(\frac{\rm{FWHM}_{H\alpha}}{10^{3}\rm{kms}^{-1}}\right)^{1.03\pm0.03}\rm{kms}^{-1}
\label{eq:HaHb}
\end{equation}

\noindent The black hole mass is then determined from the H$\beta$ FWHM using:

\begin{equation}
M_{\rm{BH}}/M_\odot=10^{6.91}\left(\frac{\rm{FWHM_{H\beta}}}{1000 \rm{kms}^{-1}}\right)^2\left(\frac{L_{5100}}{10^{44}\rm{erg s}^{-1}}\right)^{0.5}
\label{eq:BHmass}
\end{equation}

\noindent where L$_{5100}$ is estimated using the best-fit model SEDs
of each quasar. We assume that the bolometric luminosity is given by
L$_{\rm{bol}}$=7L$_{5100}$ \citep{Vestergaard:06} and use these to
calculate the Eddington fraction. The mass accretion rate is given by:

\begin{equation}
\frac{dm}{dt}=0.18 \frac{1}{\epsilon}\left(\frac{L_{bol}}{10^{46}\rm{erg s}^{-1}}\right)M_\odot \rm{yr}^{-1}
\label{eq:dmdt}
\end{equation}

\noindent where the accretion efficiency, $\epsilon$ is assumed to be
10 per cent and the factor of 0.18 results from the units being used in Eq \ref{eq:dmdt}. The derived luminosities, masses, Eddington fractions and
accretion rates are summarised in Table \ref{tab:lines}. The quoted
errors are derived by propagating the fit errors to the H$\alpha$
linewidths. 

%\begin{table*}
%\begin{center}
%\caption{Bolometric luminosities, virial black-hole masses, Eddington
%fractions and mass accretion rates derived for the 12 reddened quasars.}
%\label{tab:lines}
%\begin{tabular}{lccccc}
%\hline \hline
%Name & log$_{10}$(L$_{\rm{bol}}$/erg s$^{-1}$) & log$_{10}$(M$_{\rm{BH}}$/M$_\odot$) & L$_{\rm{bol}}$/L$_{\rm{Edd}}$ & dm/dt (M$_\odot$ yr $^{-1}$) \\
%\hline
%ULASJ0016$-$0038 & 46.56 & 9.32$\pm$0.06 & 0.13$\pm$0.05 & 6.49 \\
%ULASJ0041 & 46.86 & 9.49$\pm$1.73 & 1.72$\pm$0.31 & 13.05 \\
%ULASJ0041$-$0021 & 47.01 & 9.56$\pm$0.08 & 0.21$\pm$0.01 & 18.34 \\
%ULASJ0141+0101 & 46.59 & 9.41$\pm$0.04 & 0.16$\pm$0.02 & 7.01 \\
%ULASJ0144-0114 & 46.89 & 9.42$\pm$0.80 & 2.61$\pm$0.22 & 13.97 \\
%ULASJ0144$-$0114 & 46.99 & 9.47$\pm$0.07 & 0.26$\pm$0.02 & 17.78 \\
%ULASJ0144+0036 & 46.60 & 9.44$\pm$0.63 & 0.91$\pm$0.06 & 7.17 \\
%ULASJ0144+0036 & 46.78 & 9.53$\pm$0.03 & 0.14$\pm$0.01 & 10.82 \\
%ULASJ0221$-$0019 & 46.53 & 8.96$\pm$0.05 & 0.28$\pm$0.02 & 6.16 \\
%ULASJ1539+0557 & 47.86 & 9.87$\pm$0.07 & 0.75$\pm$0.06 & 129.51 \\
%ULASJ1002+0137 & 46.50 & 8.83$\pm$0.08 & 0.35$\pm$0.03 & 5.74 \\
%ULASJ1234+0907 & 48.20 & 10.43$\pm$0.07 & 0.46$\pm$0.04 & 286.36 \\
%ULASJ1455 & 46.71 & 8.84$\pm$1.22 & 5.70$\pm$0.79 & 9.23 \\
%ULASJ1455+1230 & 46.92 & 8.94$\pm$0.06 & 0.73$\pm$0.05 & 15.05 \\
%ULASJ2200+0056 & 47.13 & 9.18$\pm$0.05 & 0.70$\pm$0.04 & 24.519 \\
%ULASJ2224$-$0015 & 46.72 & 8.93$\pm$0.03 & 0.48$\pm$0.02 & 5.74 \\
%\hline
%\end{tabular}
%\end{center}
%\end{table*}

\begin{table*}
\begin{center}
\caption{Bolometric luminosities, virial black-hole masses, Eddington
fractions and mass accretion rates derived for the 12 reddened quasars.}
\label{tab:lines}
\begin{tabular}{lccccc}
\hline \hline
Name & log$_{10}$(L$_{\rm{bol}}$/erg s$^{-1}$) & log$_{10}$(M$_{\rm{BH}}$/M$_\odot$) & L$_{\rm{bol}}$/L$_{\rm{Edd}}$ & dm/dt (M$_\odot$ yr $^{-1}$) \\
\hline
ULASJ0016$-$0038 & 46.9 & 9.32$\pm$0.06 & 0.20$\pm$0.05 & 13 \\
%ULASJ0041 & 46.86 & 9.49$\pm$1.73 & 1.72$\pm$0.31 & 13.05 \\
ULASJ0041$-$0021 & 47.3 & 9.56$\pm$0.08 & 0.30$\pm$0.01 & 37 \\
ULASJ0141+0101 & 46.9 & 9.41$\pm$0.04 & 0.17$\pm$0.02 & 14 \\
%ULASJ0144-0114 & 46.89 & 9.42$\pm$0.80 & 2.61$\pm$0.22 & 13.97 \\
ULASJ0144$-$0114 & 47.3 & 9.47$\pm$0.07 & 0.37$\pm$0.02 & 36 \\
%ULASJ0144+0036 & 46.60 & 9.44$\pm$0.63 & 0.91$\pm$0.06 & 7.17 \\
ULASJ0144+0036 & 47.1 & 9.53$\pm$0.03 & 0.19$\pm$0.01 & 22 \\
ULASJ0221$-$0019 & 46.8 & 8.96$\pm$0.05 & 0.40$\pm$0.02 & 12 \\
ULASJ1539+0557 & 48.2 & 9.87$\pm$0.07 & 1.07$\pm$0.06 & 264 \\
ULASJ1002+0137 & 46.8 & 8.83$\pm$0.08 & 0.50$\pm$0.03 & 11 \\
ULASJ1234+0907 & 48.5 & 10.43$\pm$0.07 & 0.65$\pm$0.04 & 580 \\
%ULASJ1455 & 46.71 & 8.84$\pm$1.22 & 5.70$\pm$0.79 & 9.23 \\
ULASJ1455+1230 & 47.2 & 8.94$\pm$0.06 & 1.04$\pm$0.05 & 30\\
ULASJ2200+0056 & 47.4 & 9.18$\pm$0.05 & 0.99$\pm$0.04 & 50 \\
ULASJ2224$-$0015 & 47.0 & 8.93$\pm$0.03 & 0.68$\pm$0.02 & 19 \\
\hline
\end{tabular}
\end{center}
\end{table*}

In addition to these fit-errors, there are also systematic errors affecting
the black-hole mass estimates, that should be discussed. We therefore
highlight several caveats relevant to the black-hole mass estimates
in Table \ref{tab:lines}, which may result in
large systematic uncertainties. In all cases,
black-hole masses were estimated using the FWHM of a single Gaussian
fit to the H$\alpha$ line profile. However, using the broad component
only to estimate the black-hole masses would result in an increase in
the black-hole masses and correspondingly a decrease in the Eddington
fractions by up to a factor of $\sim$2. Furthermore, we note that
several of our Gaussian line profiles in Fig.~\ref{fig:Hagauss2} and
Fig.~\ref{fig:Hagauss} are highly asymmetric showing large velocity
shifts between the different components.

\subsubsection{Evidence for Outflows:}

\label{sec:outflows}

In most cases, the broad component shows a clear blueshift relative to
the intermediate component. This blueshift is most pronounced in
ULASJ0016-0038 and ULASJ0041-0021 where the broad component has a velocity of
--1400\,kms$^{-1}$ and --800\,kms$^{-1}$ relative to the intermediate
component in the two respective sources. In ULASJ1455+1230 however, the
broad component is redshifted by +760\,kms$^{-1}$ relative to the
intermediate component. Outflowing material in a bipolar flow can produce both blue and red-shifted emission so these velocity shifts indicate the presence of
strong outflows that are affecting the broad line region of
the quasar. Signatures of outflows, while not generally seen in the UV-luminous quasar population, are not uncommon in samples of red BAL quasars where they have been shown to quench star formation within their host galaxies \citep{Farrah:12}. These flows can broaden the H$\alpha$
linewidths leading to an over-estimation of the virial black-hole
masses and consequently an under-estimate of the Eddington
fractions. We do not correct for this effect in our black-hole mass estimates but caution that excluding the blue/red asymmetric component from the H$\alpha$ line fits in quasars showing evidence for outflows, results in a decrease in the estimated black-hole masses by a factor of $\sim$2--3. 

\subsection{Notes on Individual Sources}

\label{sec:notes}

Two of the sources in our reddened AGN sample are particularly
interesting and merit further discussion: \\

\textit{\underline{ULASJ1234+0907}}: ULASJ1234+0907 is our reddest quasar with
$(H-K)$=2.3 and an inferred dust extinction of A$_V$=6.0 derived from this colour. The
extinction corrected bolometric luminosity for this source is
L$_{\rm{bol}}>$10$^{48}$erg s$^{-1}$ and the inferred black-hole mass
is $>$10$^{10}$M$_\odot$ from the very broad H$\alpha$ line seen in
Fig.~\ref{fig:spectra}. This makes ULASJ1234+0907 one of the most
bolometrically luminous and massive black-holes known at $z\sim$2. As
discussed in detail in Section~\ref{sec:disc}, the extinction-corrected
absolute magnitude for ULASJ1234+0907 is M$_i \sim -$31--$-$32 which is comparable to
the known lensed reddened quasar PKS0132 at similar redshifts
\citep{Hall:02, Gregg:02}. Our ongoing follow-up observations of
ULASJ1234+0907 at long wavelengths will enable a detailed characterisation
of its host galaxy properties, shedding light on the reason for the
extreme luminosity and dust extinction. Although we cannot exclude
lensing as a possible reason for the extreme luminosity of this source
we discuss why this is statistically improbable in
Section~\ref{sec:lens}. \\

\textit{\underline{ULASJ1002+0137}}: ULASJ1002+0137 is in the COSMOS field and
and has previously been studied by \citet{Mainieri:07} in their
analysis of X-ray AGN from the XMM-Newton Wide Field Survey. The
optical counterpart to the X-ray source is presented in
\citet{Brusa:10}. Although a redshift of $z=$0.784 was initially
derived by \citet{Mainieri:07} and \citet{Trump:09}, the optical
spectroscopic redshift of $z=$1.592 in \citet{Brusa:10} agrees well
with our H$\alpha$ redshift and the photometric redshift of
\citet{Salvato:09}. The X-ray spectrum is best fit by an absorbed
power-law and an Fe K$\alpha$ line at $z=$0.784. The multi-wavelength
analysis and optical spectroscopy produces a classification as a
Narrow-Line/Type 2 AGN \citep{Brusa:10}. Our measured H$\alpha$ FWHM
of $\simeq$3000\,kms$^{-1}$ would suggest that the broad-line region
is at least partially visible. However, this FWHM is significantly
smaller than the broad components for the rest of the sample and
ULASJ1002+0137 is the only source where a single Gaussian provides an
adequate fit. If the broad-line region is partially obscured by a
dusty torus, the large dust extinction of A$_V$=3.2 inferred for this
source would mean that the broad-line flux is expected to be more
suppressed in the rest-frame ultraviolet than in the optical. This may
have prevented this source from being classified as a broad-line AGN
from the optical spectrum.

\begin{figure}
\begin{center}
\includegraphics[width=6.5cm,height=5.0cm,angle=0]{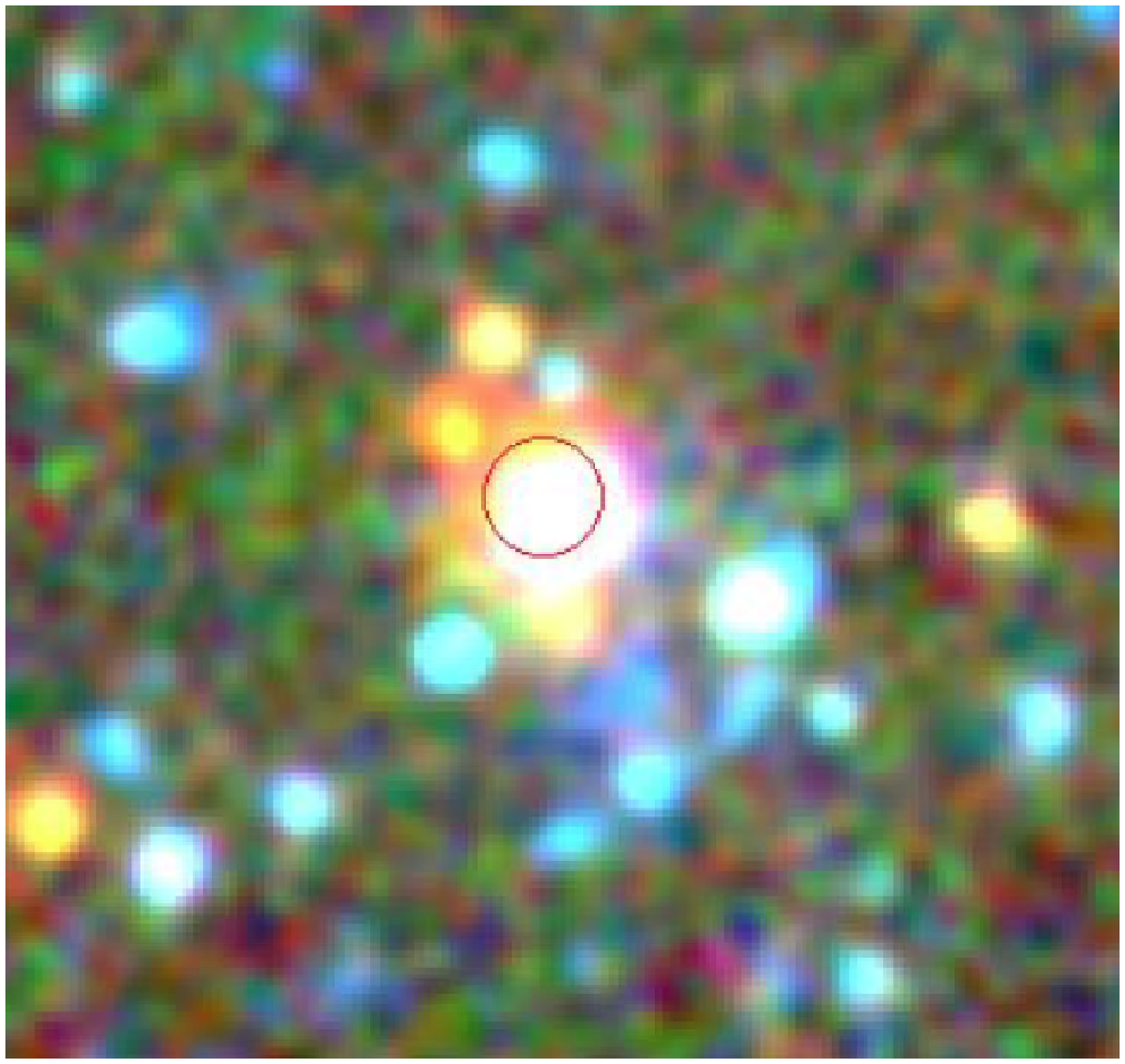}
\caption{COSMOS $B$, $i$, $K$ composite 15$\times$15$^{\prime \prime}$
cut-out around ULASJ1002+0137. The quasar is marked with the circle. An
excess of blue galaxies is seen around the source and the three
sources located southwest of the quasar could be a lensed arc.}
\label{fig:1002_2}
\end{center}
\end{figure}

We should also consider the possibility that this is a lensed quasar
with the source at $z=$1.595 and the lens at $z=$0.784. In Fig.
\ref{fig:1002_2} we show a 15$\times$15$^{\prime\prime}$ COSMOS $BiK$
composite around the source at a resolution of 0.15$^{\prime\prime}$
per pixel \citep{McCracken:10}. An excess of blue sources is seen
around the red quasar and the three blue sources to the southwest of
ULASJ1002+0137 could be a lensed arc although the suggestion is highly
tentative. We discuss the possibility of lensing in our sample further
in Section~\ref{sec:lens}.

\section{DISCUSSION}
\label{sec:disc}

\subsection{Lensing in the Reddened Quasar Population}

\label{sec:lens}

\begin{figure}
\begin{center}
\includegraphics[width=8.5cm,height=6.0cm,angle=0]{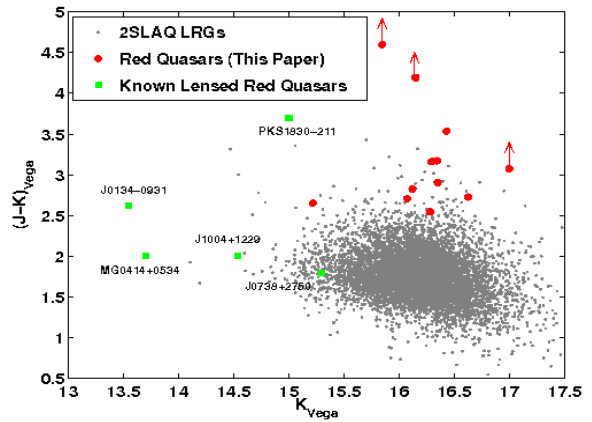}
\caption{$(J-K)$ versus $K$ (Vega) colour-magnitude diagram comparing our sample of
reddened quasars to some well-known lensed reddened quasars at similar
redshifts as well as lower redshift LRGs, representative of the likely
lens. Our sample of reddened quasars are always fainter and almost
always redder than the known lensed reddened quasars and significantly
redder than the LRG lenses.}
\label{fig:JKlens}
\end{center}
\end{figure}

It has previously been demonstrated that lensing may be responsible for
reddening the optical and infra-red colours of quasars
\citep{Malhotra:97}. There are various explanations for why this may
be the case. Firstly, dust from a lens galaxy could extinct the light
coming from the quasar as in the case of the lensed quasar PKS1830-211
\citep{Courbin:98}. Alternatively, there could be a contribution to the 
quasar colours from galactic starlight within the lensing galaxy. There are various examples of
lensed quasars in the literature which were selected based on their
infra-red excess. These include J0134-0931 otherwise known as PKS0132
\citep{Winn:02, Gregg:02, Hall:02}, FIRSTJ1004+1229 \citep{Lacy:02} and
SDSSJ1313+5151 \citep{Ofek:07}. In Fig.~\ref{fig:JKlens} we compare the
colours and magnitudes of our reddened quasars to some of these known
lensed quasars, selected based on their infra-red excess. In the same
plot we also show the distribution of LRGs from the 2SLAQ survey at
0.4$<z<$0.8 \citep{Cannon:06} which have been matched to the UKIDSS-LAS using a
matching radius of 1.5$^{\prime\prime}$.  These LRGs may be considered
to be representative of the likely lens population.

In the previous section we presented some very tentative evidence for
ULASJ1002+0137 being a lensed object based on the two different redshifts
measured and the excess of blue galaxies in the deep COSMOS images. It
is interesting to consider the likelihood of lensing in our population of NIR selected red quasars. Based on Fig.~\ref{fig:JKlens} it can be seen that our
reddened quasars are both apparently fainter and redder in colour than
most of the known lensed red quasars. PKS1830-211 has a very red
observed colour on account of lying in the Galactic plane but this is
not generally true of the other lensed objects. Lensing galaxies are
typically LRGs at $z<$1 in which case they should be detectable in the
SDSS $i$ and $z$ bands as well as the bluer UKIDSS bands. The reddened
quasars that are undetected in the SDSS as well as the UKIDSS $Y$ and
$J$-bands include our reddest and intrinsically brightest source
ULASJ1234+0907. Even if the lensing galaxy had a redshift as high as
$z\simeq$1, we would expect to be able to detect it in the UKIDSS $Y$
and $J$-bands. 

We conclude that although we cannot rule out the
possibility of some of the reddened quasars in our sample being
lensed, in general this is unlikely to be the case for the bulk of the
population and dust extinction within the quasar host is a more
plausible explanation for the extreme colours observed in these
sources.

\subsection{Reddened Quasars: A Transition Population Between the Starburst \& Optical AGN Phases?}

Our study of extremely red near infra-red selected bright quasars has
unearthed a population of broad-line AGN at $z\sim$2 corresponding to
the main epoch of galaxy formation with significant dust extinction of A$_V \gtrsim 2$mags and comparable to that measured in the
sub-millimetre galaxy population \citep{Takata:06}. We have already
ruled out lensing as a possible explanation for the extreme colours in
the bulk of this population. The very large observed widths ($\sim$5000 km/s) of the
Balmer emission lines support the hypothesis that the reddening is
produced by dust within the host galaxy rather than a molecular torus
which would obscure our view of the AGN broad-line region. An
explanation that does not rely on viewing orientation is supported by the
fact that the bolometric luminosities are typically very high and comparable
to some of the most luminous optically selected QSOs. The inferred mass 
accretion rates are therefore also high. The fuel for this
black-hole accretion likely comes from intense star formation within the
host galaxy consistent with the high dust extinctions being produced
by a major starburst event. Highly star-forming, dust-obscured,
ultra-luminous infra-red galaxies (ULIRGs) and quasars are thought to
be different observational manifestations of the same phenomena
\citep{Sanders:88}.  Major-merger induced starbursts appear as ULIRGs
or SMGs at $z\sim$2 \citep{Blain:02, Smail:02} and as the dust clears
from the decaying starburst, the central nuclear region is revealed as
an optically bright quasar. Our sample of infra-red bright reddened
quasars may therefore represent a transition phase between the
starburst and optical quasar phases where the dust has not yet fully
cleared preventing the sources from being selected as
ultraviolet-luminous quasars. This \textit{blowout} phase is also 
evidenced by the presence of strong outflows that seem to be affecting
the H$\alpha$ line profiles in some of our quasars.

\begin{figure*}
\begin{center}
\begin{minipage}[c]{1.00\textwidth}
\centering
\includegraphics[width=8.5cm,height=6cm,angle=0] {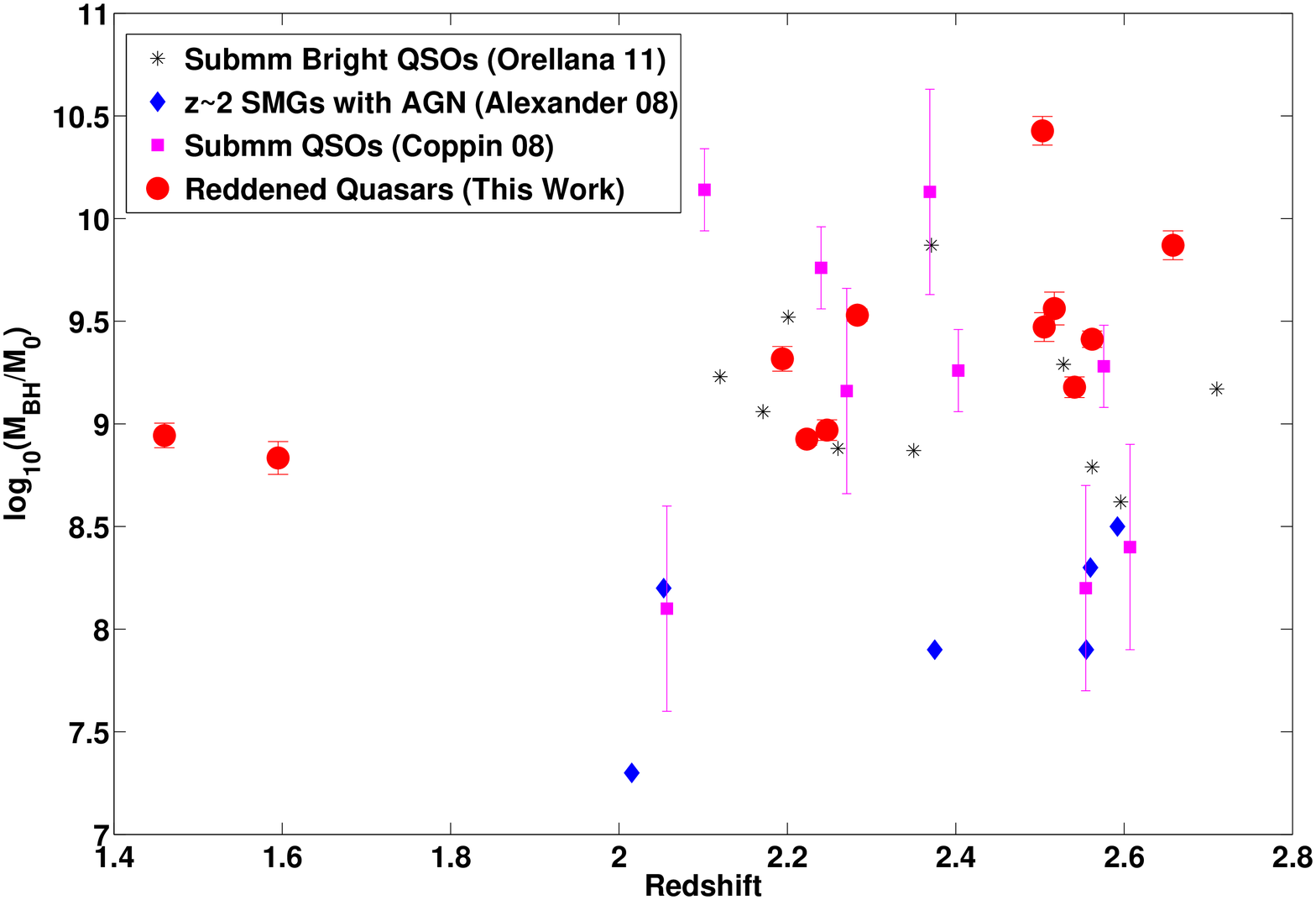}
\includegraphics[width=8.5cm,height=6cm,angle=0] {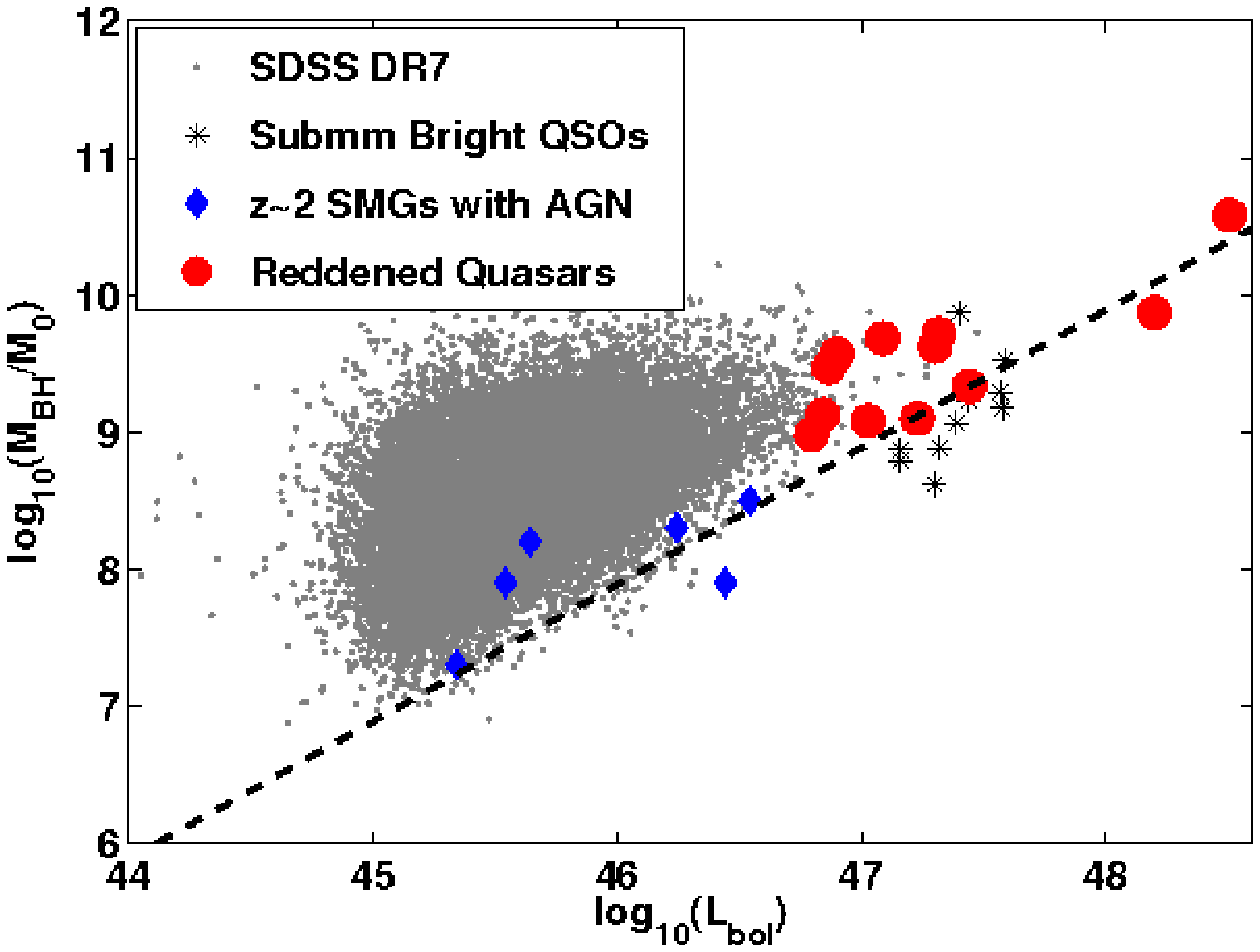}
\end{minipage}
\caption{\textit{Left:} Redshift versus black-hole mass of our
  reddened quasars compared to sub-millimetre bright quasars from
  \citet{Coppin:08} and \citet{Orellana:11} as well as X-ray selected
  sub-millimetre galaxies hosting AGN from \citet{Alexander:08}. Our
  sample of reddened quasars constitute some of the most massive black
  holes at $z\sim$2. \textit{Right:} Bolometric luminosity versus black-hole mass
  for our sample of reddened quasars compared to the
  optically selected quasars from the SDSS \citep{Shen:11},
  sub-millimetre bright quasars \citep{Orellana:11} and sub-millimetre
  galaxies hosting AGN \citep{Alexander:08}. The dashed line corresponds
  to Eddington accretion. Note, the scaling relations used to derive the
  black-hole masses and bolometric luminosities means the two quantities are
  necessarily correlated.}
\label{fig:MBH}
\end{center}
\end{figure*}

We now compare our sample properties to those of AGN samples selected
at other wavelengths. These include the optically selected quasar
population from SDSS DR7 \citep{Shen:11}, sub-millimetre bright
quasars from \citet{Coppin:08} and \citet{Orellana:11} and a subset of
sub-millimetre galaxies hosting AGN that were selected from X-ray
studies \citep{Alexander:08}. The sub-millimetre quasars in
\citet{Orellana:11} are all optically bright blue quasars whereas
those in \citet{Coppin:08} include both standard blue optical quasars
as well as redder X-ray absorbed quasars and quasars detected in
blank-field sub-millimetre surveys. In Fig. ~\ref{fig:MBH} we plot the
black-hole masses as a function of redshift as well as bolometric 
luminosity for all these samples. The AGN bolometric luminosity for 
the SMGs was estimated by scaling the X-ray luminosities by a factor of 35 \citep
{Alexander:08}. Note, that the scaling relations we use implicitly assume that 
the black-hole mass correlates with the bolometric luminosity so the trend seen 
in Fig.~\ref{fig:MBH} should be interpreted with caution.

Fig.~\ref{fig:MBH} shows that our sample of reddened quasars
constitute some of the most massive AGN at $z\sim$2 with black-hole
masses comparable to the quasars studied by \citet{Coppin:08} and
\citet{Orellana:11} and an order of magnitude larger than SMGs hosting
AGN at similar redshifts. We note in particular that the most massive
black-holes in the \citet{Coppin:08} sample are the optically bright
subset of their population and the X-ray absorbed and sub-millimetre
detected sources in their sample have relatively modest black-hole
masses. Our quasars are therefore as massive as the most luminous
optically selected quasars but still have significant amounts of
dust. These results are consistent with a scenario where the growth of
the black-hole follows the starburst phase. In other words, the
quasars found in blank sub-millimetre surveys and the SMGs hosting AGN
have not yet had time to significantly grow their black holes and
represent an earlier evolutionary stage in the lifecycle of these
sources. 

Recently \citet{Hickox:12} used clustering arguments to 
provide indirect evidence that powerful starbursts and optical quasars
occur in the same systems. Our ongoing follow-up of the reddened quasars at submillimetre 
wavelengths would provide direct evidence for this link between starbursts and AGN. If
the reddened quasars are indeed transitioning from starbursts to UV-bright quasars, 
the expectation is that the host galaxies of our population are likely to be more 
luminous at far infra-red and submillimetre wavelengths
than the hosts of the optical QSO population (e.g. \citet{Isaak:02} and \citet{Priddey:03}). 
If we 
assume that most of the mass in quasars is assembled at $z>=$ 2 
\citep{McLure:04} consistent with \textit{downsizing} in the quasar 
population, then the reddened quasars are likely being observed at 
a younger stage than UV luminous quasars of similar mass, which have 
presumably already been through an obscured phase associated with the 
bulk of their mass assembly. 

New large-area far infra-red to millimetre surveys such as $\textit{Herschel}$-ATLAS and
surveys with the South Pole Telescope, could offer the opportunity to directly detect 
the host galaxies of this rare population due to surveying much larger areas than covered 
by previous surveys of the SMG population.  

\subsection{Obscured Fraction of Type 1 AGN}

\label{sec:frac1}

We now attempt to derive some constraints on the space density of
obscured Type 1 AGN in order to address whether these optically
obscured sources constitute a significant fraction of the total Type 1
AGN population. Previous studies of highly obscured broad-line quasars have relied on matching to radio surveys (e.g. \citet{Webster:95, Glikman:07}) so it is interesting to consider whether similar conclusions about the obscured fraction are reached by studying the more numerous radio-quiet quasars. Obscured, in the context of the following discussion, is taken to mean quasars with A$_V \gtrsim 2$.

\begin{figure*}
\begin{center}
\begin{minipage}[c]{1.00\textwidth}
\centering
\includegraphics[width=8.5cm,height=6cm,angle=0] {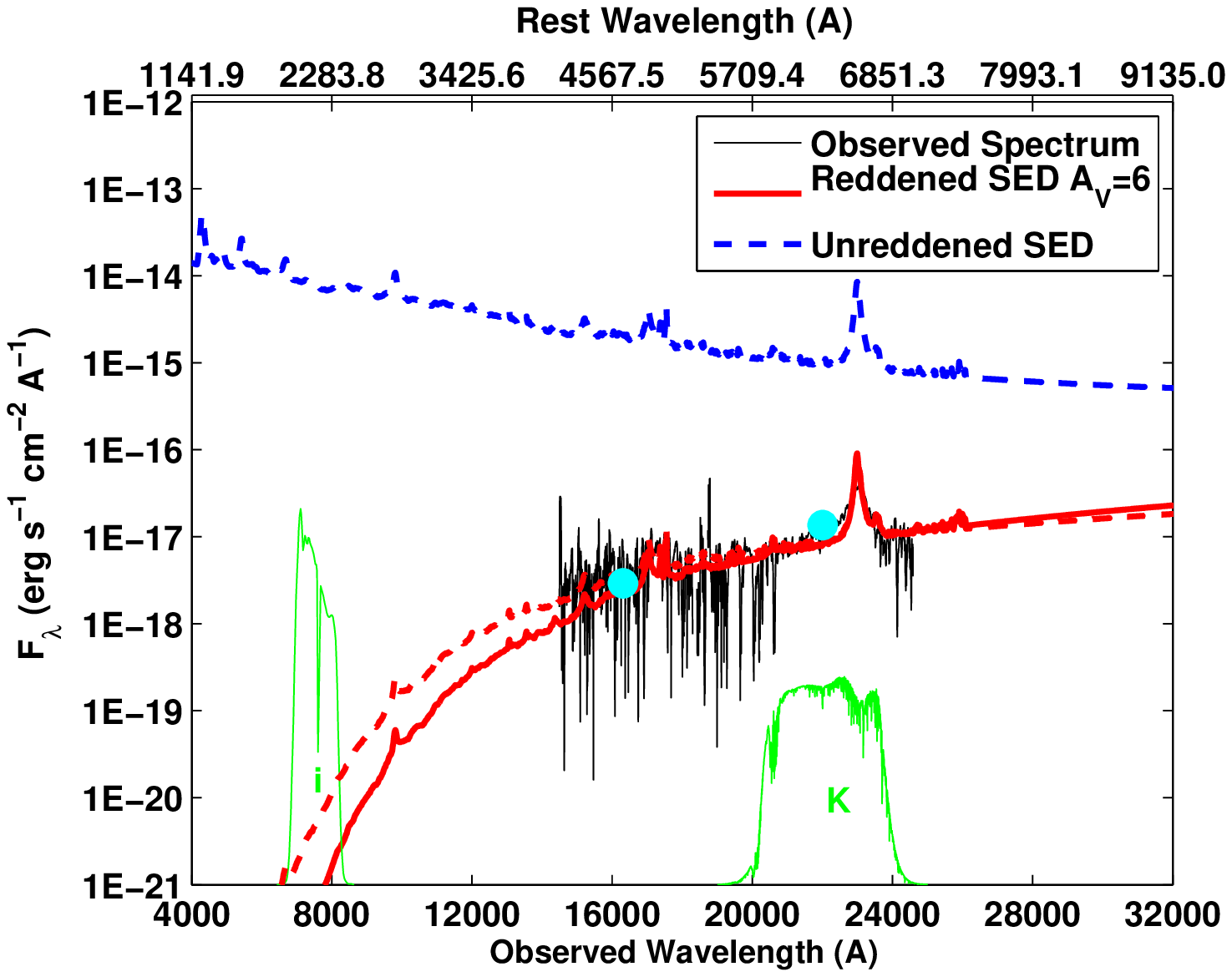}
\includegraphics[width=8.5cm,height=6cm,angle=0] {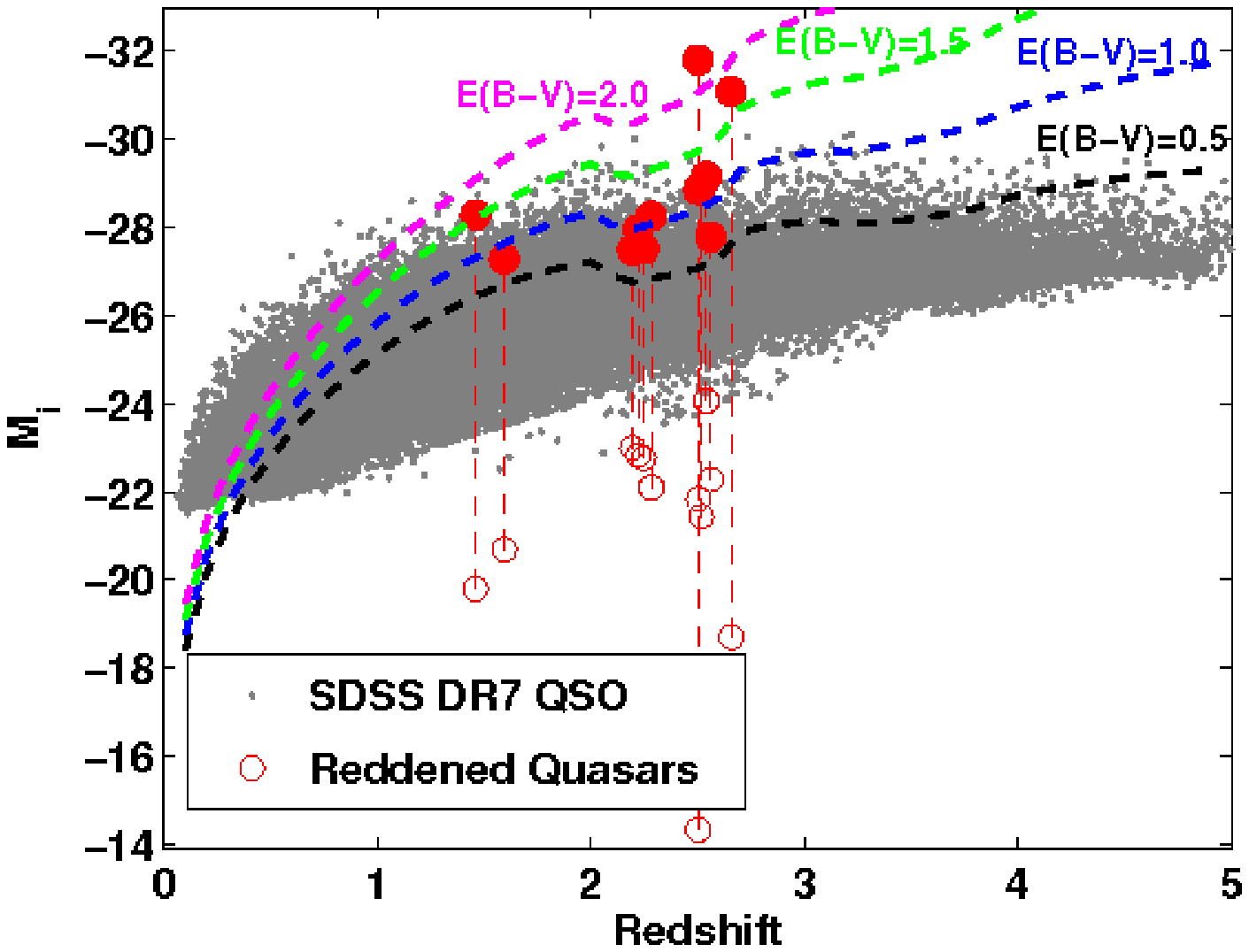}
\end{minipage}
\caption{\textit{Left:} Observed VLT-SINFONI spectrum for ULASJ1234+0907
  (black) compared to the best-fit reddened model SED (red) with
  A$_V$=6.05 and the unreddened SED (blue) which has been corrected
  for extinction using the adopted reddening law
  (Section~\ref{sec:sed}). We also plot a model with A$_V$=4.96 as the red dashed line. This model produces a slightly better match to the spectrum continuum than the A$_V$=6.05 model fit to the photometry. The UKIDSS $H$ and $K$-band fluxes that have been used to derive the higher $A_V$ value are shown as the filled circles. The optical SDSS $i$-band bandpass as well
  as the near infra-red ULAS $K$-band bandpass are also shown for
  reference. ULASJ1234+0907 is a very extreme object with an effective
  reddening of 17.5 mags in the SDSS $i$-band ($\simeq$2250\,\AA \ in
  the rest-frame) making it one of the most intrinsically luminous
  quasars known. \textit{Right:} Absolute $i$-band magnitude versus
  redshift for our reddened quasars compared to the optical quasar
  population from SDSS DR7. The open circles correspond to the
  reddened absolute magnitudes and the filled circles represent the
  extinction corrected absolute magnitudes with the red dashed lines
  indicating the effect of extinction for each object. On correcting
  for the large extinction values, the reddened quasars mostly lie at
  the most luminous end of the optical sample with no
  spectroscopically confirmed SDSS quasars that are intrinsically as
  bright as ULASJ1234+0907 and ULASJ1539+0557. The plot also shows the
  tracks of a $K=17$ quasar with different values of $E(B-V)$ in the $M_i--z$ plane corresponding to the
  flux-limit of our survey.}
\label{fig:Mi}
\end{center}
\end{figure*}

In Section~\ref{sec:sed} we fitted model SEDs to the
broadband colours of our reddened quasars in order to infer a dust
extinction for each quasar. Using this dust extinction value and our
reddening law (Section~\ref{sec:sed}), we correct each model SED in
order to derive an unreddened SED for each quasar. The left-panel of
Fig.~\ref{fig:Mi} shows an example of this procedure for our reddest
quasar, ULASJ1234+0907. It can immediately be seen that the extinction in
the SDSS optical bands is significant; an A$_V$=6.0 corresponds to an
extinction of 17.5 magnitudes at rest-frame 2250\,\AA\@ which makes the quasar undetectable in even deep ground-based optical surveys. Correcting for
the extinction makes ULASJ1234+0907 one of the brightest quasars known at
these redshifts with an absolute magnitude of M$_i$=-31.8. In Fig.~\ref{fig:Mi} we also plot a model with A$_V$=4.96 for ULASJ1234+0907 which produces a slightly better match to the spectrum continuum. Adopting this lower extinction value rather than the one inferred from fitting to the broadband photometry, still makes this source a very luminous quasar with M$_i$=$-$31.0 and changes our conclusions little. In Banerji
et al. (2012, in preparation) we present follow-up observations of
this source at multiple wavelengths in order to better characterise
its nature.

In the right panel of Fig.~\ref{fig:Mi}, we plot the extinction
corrected absolute magnitudes versus redshift for all our reddened
quasars compared to the SDSS optically selected population of
\citet{Shen:11}. The total reddened quasar sample from Paper I and this work covers an area of 955\,deg$^2$. We also list these extinction corrected $i$-band absolute magnitudes in Table \ref{tab:Mi}. We have used a correction of +0.596
\citep{Richards:06} to convert between M$_i$[z=2] and M$_i$ for the
SDSS quasars. The plot illustrates both the reddened absolute
magnitudes as well as the de-reddened absolute magnitudes of our sample in order to
demonstrate the effect of extinction on the inferred optical
brightness. As can be seen from the figure, the extincted optical magnitudes lie well 
below the SDSS flux limit for this sample but correcting for this dust 
extinction makes the reddened quasars as luminous as some of the brightest
SDSS sources and in some cases, brighter than any known SDSS quasar. This is to be expected given that we are selecting red quasars in a NIR flux-limited sample. Consequently, for a constant NIR flux, more heavily obscured sources are going to be more intrinsically luminous. This can be seen in the right-panel of Fig.~\ref{fig:Mi} which also shows our $K$-band selection boundary in the M$_i-z$ plane. Our $(J-K)>2.5$ colour selection broadly corresponds to selecting quasars at z$\sim$2 with $E(B-V) \gtrsim 0.5$ (Paper I), albeit with some scatter. We note however that while the NIR selection allows us to detect the most heavily obscured intrinsically luminous quasars at $z\sim$2, it is biased against heavily obscured quasars with more modest luminosities at the same redshifts - i.e. a quasar with $E(B-V)=1.5$ for example, would have to be one of the most luminous quasars known to make it into our K-band sample at $z\sim$2. We discuss this selection effect further later in this Section.

\begin{table}
\begin{center}
\caption{De-reddened absolute $i$-band magnitudes for all our spectroscopically confirmed red quasars. The quasars marked with an asterisk extend to a slightly fainter flux-limit of K$_{AB}<$18.9. They were targetted due to the presence of ancillary multi-wavelength information but do not represent a complete sample and are therefore excluded from the space-density calculations.}
\label{tab:Mi}
\begin{tabular}{lc}
\hline \hline
Name & Dereddened M$_i$ \\
\hline
ULASJ0016$-$0038 &$-$27.49 \\
ULASJ0041$-$0021 & $-$28.94 \\
*ULASJ0141+0101 & $-$27.78\\
ULASJ0144+0036 & $-$28.26 \\
ULASJ0144$-$0114 & $-$28.85\\
ULASJ0221$-$0019 & $-$27.52 \\
ULASJ1539+0557 & $-$31.08 \\
*ULASJ1002+0137 & $-$27.29 \\
ULASJ1234+0907 & $-$31.81 \\
ULASJ1455+1230 & $-$28.29 \\
ULASJ2200+0056 & $-$29.16 \\
ULASJ2224$-$0015 & $-$27.96 \\
\hline
\end{tabular}
\end{center}
\end{table}

Fig.~\ref{fig:Mi} clearly demonstrates that the reddened quasars we have found are
some of the most intrinsically luminous quasars known at z$\sim$2.  How does their
space density compare to the optically selected population? Estimates of the
space density of optical quasars in the SDSS for example are complicated by the fact
that the SDSS is highly incomplete in the redshift range 2.2$<z<$2.7, due to the similarity of
Galactic star and quasar colours at these redshifts. \citet{Richards:02, Richards:06} have however
studied this incompleteness in detail and corrected for it when deriving the quasar luminosity function.
The selection functions derived by these authors were based on comparisons to simulations but have 
recently been verified empirically using a large sample of KX-selected quasars \citep{Maddox:12}. 
We can therefore use the SDSS luminosity function to predict the number of UV-bright quasars expected 
in the redshift range 1.2$<$z$<$1.75 or 2.0$<$z$<$2.7 over our survey area of 955\,deg$^2$. These numbers of UV-bright
quasars should then be directly comparable to the numbers of NIR bright reddened quasars in our sample. 
We note here that our sample extends to brighter absolute magnitudes
than the SDSS but the parametric form of the luminosity function can reasonably be extrapolated to these 
bright magnitudes in order to provide an estimate for the number of very luminous UV-bright quasars. The 
completeness corrected total number of UV-bright quasars at 1.2$<$z$<$1.75 or 2.0$<$z$<$2.7 over 955 deg$^2$ of sky 
predicted from the SDSS luminosity function, is given in Table \ref{tab:comp}.  

Our sample of reddened quasars also suffers from incompleteness. The first
source of incompleteness arises because not all the red quasar
candidates in our UKIDSS sample were targetted for spectroscopy. There
are six candidates with colours consistent with identifications as
reddened quasars but without spectra. Best-fit $E(B-V)$ values for
these sources are calculated, just as for the spectroscopically
confirmed sample, but assuming redshifts of 1.5 and 2.5. Absolute
$i$-band magnitudes can then be calculated as before. The results of this
exercise are summarised in Table \ref{tab:nospec}. It can be seen that the absolute $i$-band
magnitudes are similar regardless of the redshift. However, the red
$(H-K)$ colours of these sources strongly favour the higher redshift
assumption.

\begin{table*}
\begin{center}
\caption{Six quasar candidates for which no spectra were obtained
  along with their best-fit $E(B-V)$ values assuming $z=$1.5 or $z=$2.5 as
  well as de-reddened absolute $i$-band magnitudes corresponding to
  these two redshifts.}
\label{tab:nospec}
\begin{tabular}{lcccc}
\hline \hline
Name & E(B-V),z=1.5 & M$_{i}$, z=1.5 & E(B-V), z=2.5 & M$_{i}$, z=2.5 \\
\hline
ULASJ0946+0016 & 1.38 & $-$28.82 & 0.80 & $-$28.59 \\
ULASJ1200+0423 & 1.65 & $-$29.29 & 0.98 & $-$29.05 \\
ULASJ1320+0948 & 1.22 & $-$28.36 & 0.69 & $-$28.12 \\
ULASJ0121+0107 & 1.06 & $-$28.22 & 0.60 & $-$28.04 \\
ULASJ0204+0106 & 1.18 & $-$28.34 & 0.67 & $-$28.13 \\
ULASJ2346$-$0038 & 1.31 & $-$29.62 & 0.75 & $-$29.38 \\
\hline
\end{tabular}
\end{center}
\end{table*}

The second source of incompleteness is due to morphological
misclassification of candidates, whereby some reddened quasars,
classified as galaxies in the $K$-band, are not included in our
sample. In Paper I we have shown that at the brighter
end of our survey the maximum morphological incompleteness in our redshift
range is $\sim$20 per cent. 

Finally, as already pointed out in Fig.~\ref{fig:Mi}, our flux-limited NIR sample only allows selection of the most intrinsically luminous quasars at $z\sim$2 with large amounts of dust extinction. More modest luminosity quasars with similar or higher levels of dust obscuration require a deeper $K$-band survey to be found so we could potentially be missing large numbers of highly reddened quasars at modest luminosities. We estimate our survey incompleteness to a typical A$_V$=2.5 quasar by assuming that every SDSS quasar at $2.0<z<2.5$ for example, has this level of dust extinction, and counting the fraction of these dusty quasars that satisfy our survey selection criteria ($K<16.5$ and $(J-K)>2.5$; Vega) in each of the absolute magnitude bins in Table \ref{tab:comp}. We find that our survey would recover $>$99\% of A$_V$=2.5 quasars with $-32<M_i<-30$ at these redshifts with the completeness dropping to $\sim$90\%, 60\% and 10\% in the subsequent luminosity bins. We correct the numbers of quasars in our survey for all these sources of incompleteness to derive the upper limits quoted in Table \ref{tab:comp}. These are then compared to the numbers of UV-bright quasars from the SDSS luminosity function over the same redshift range and area, to derive an obscured fraction in each absolute magnitude bin.

\begin{table*}
\begin{center}
\caption{Number of quasars at 1.2$<$z$<$1.75 or 2.0$<$z$<$2.7 over 955 deg$^2$ in bins of M$_i$ for our survey of red quasars compared to the SDSS DR7 sample of blue quasars (corrected for completeness). The obscured fraction represents the fraction of reddened quasars relative to the total number of quasars in a particular absolute magnitude bin. In the case of our survey, the lower limits are computed from the raw counts of reddened quasars in this work while the upper limits take into account the various sources of incompleteness in the sample detailed in Section.~\ref{sec:frac1}.}
\label{tab:comp}
\begin{tabular}{ccccc}
\hline \hline
M$_i$ & N/955deg$^2$(SDSS DR7) & N/955deg$^2$(This Work) & Obscured Fraction & N/955deg$^2$(F2M (Glikman07)) \\
\hline
--32 to --30 & 0.36 & 2--3 & 80--90\% & 1 \\
--30 to --29 & 3.3 & 1--4 & 20--60\% & 0.7 \\
--29 to --28 & 33 & 4--17 & 10--30\% & 1.8 \\
--28 to --27 & 304 & 3--50 & 1--10\% & 1 \\
\hline
\end{tabular}
\end{center}
\end{table*}

\subsection{Is the Obscured Fraction Luminosity Dependent?}

\label{sec:frac2}

Taken at face-value, Table \ref{tab:comp} seems to suggest that the obscured fraction of broad-line quasars declines with intrinsic luminosity with these red quasars only making up a relatively small fraction of modest luminosity Type 1 quasars at $z\sim$2. At very high luminosities however, heavily reddened optically faint quasars dominate the population although with only small numbers at present to exhibit this trend. Is this trend physical or can it simply be explained by selection effects? While it is true that our survey is too bright to detect heavily reddened quasars with A$_V \gtrsim 3$ at M$_i \gtrsim -28$, should large populations of such sources exist, there are several reasons to suggest that this is probably not the case. Flat-spectrum radio-source samples with very high quasar redshift completeness (e.g. \citet{Jackson:02}) indicate that the quasar population as a whole cannot consist of a very large fraction of obscured sources \citep{Ellison:01}. Although our sample size is still small, we have carefully taken into account the various selection effects in our survey and we probe down to $K$-band depths sufficient to detect most A$_V$=2.5 quasars out to M$_i<-28$ at z$\sim$2. We conclude that highly reddened quasars are likely to only dominate the counts of Type 1 quasars at the brightest absolute magnitudes, where NIR surveys may be more complete than optical surveys.

The above arguments while still tentative due to the small numbers of quasars in our sample seem plausible based on physical arguments. The duty cycle for the obscured phase is likely to be longer for the most luminous quasars, consistent with a longer time required to assemble more massive black holes. Seen another way, the obscured phase is more likely to be associated with the most intrinsically luminous quasars with high mass accretion rates. These rapidly accreting, massive, obscured black-holes should preferentially lie at z$\sim$2 corresponding to the main epoch of black-hole formation. Indeed, our two reddest and most luminous quasars ULASJ1234+0907 and ULASJ1539+0557 lie at the high redshift end of our sample at $z\sim$2.5. Our selection criteria do not in any way discriminate against similarly luminous highly reddened quasars at 1.2$<z<$1.75 but both the quasars observed at these lower redshifts have comparably modest luminosities. Again, we caution against drawing strong conclusions from the very small numbers of sources in our sample but the fact that our most reddened and most luminous quasars are concentrated at the high redshift end of our sample, is also consistent with the most massive black holes assembling the bulk of their mass at early times via an obscured phase. 

Taken together, all these facts suggest that the obscured fraction may depend on quasar luminosity with a larger proportion of the most luminous quasars being preferentially observed in an obscured phase at z$\sim$2. Our sample is still however highly incomplete for obscured quasars at M$_i \gtrsim -28$ and a deeper $K$-band sample is clearly needed to fully test this theory.

\subsection{Comparison to the 2MASS Sample}

Having demonstrated that our population of intrinsically luminous dust-reddened quasars at z$\sim$2 with $E(B-V)>0.5$ seem to represent a sizeable fraction of the total quasar population at very bright absolute magnitudes, we now make some explicit comparisons to previous studies of such objects using the shallower ($K_{\rm{Vega}}<15$) 2MASS survey \citep{Glikman:07}. Our survey represents a significant increase in depth compared to this previous work and extends 1.5--2 mags fainter in the $K$-band. Isolating only those quasars in the 2MASS survey that overlap the redshift range of our sample (1.2$<z<$1.75 or 2.0$<z<$2.7), there are only 13 reddened broad-line quasars over 2716 deg$^2$ in the \citet{Glikman:07} sample. For each of these 13 QSOs, we use the $K$-band magnitudes and dust extinctions given in \citet{Glikman:07} along with the average ($i-K$) colour of our quasar models ($\S$ \ref{sec:sed}) at that redshift, to calculate an absolute $i$-band magnitude. The counts are then scaled to our survey area of 955 deg$^2$ and once again quoted in Table \ref{tab:comp}. Table \ref{tab:comp} demonstrates that our NIR selection uncovers much larger numbers of quasars at z$\sim$2 than \citet{Glikman:07} but with similar intrinsic luminosities and much higher levels of dust-reddening. The z$\sim$2 2MASS sample has a median $E(B-V)=0.4$ while our sample over the same redshift range has a median $E(B-V)=0.8$.

We have also shown that the most heavily dust-reddened quasars in our sample lie at the highest redshifts whereas the reddest quasars in \citet{Glikman:07} are preferentially at the low redshift end of their sample due to the shallow $K$-band limit. If red quasars in which the red colours are attributed to dust extinction, represent a young phase in quasar evolution, they should preferentially be seen at z$\sim$2 corresponding to the peak of star formation and black-hole accretion. This is what is observed in our sample and suggests that other factors apart from dust extinction in the quasar host may be responsible for the red colours in the reddest quasars in the \citet{Glikman:07} sample. Selecting only quasars with $E(B-V)>$0.5 from the \citet{Glikman:07} sample in order to overlap the typical reddening in the quasars in our sample, we find a median redshift of 0.6 for these quasars. At these redshifts, the host galaxy contribution to the quasar can be significant, affecting the quasar colours. In addition, at these lower redshifts, the near infra-red $J$ and $K$-bands are also tracing hot dust emission. In a forthcoming paper, we demonstrate using \textit{WISE} photometry for our quasars that the temperature of this hot dust can significantly affect the colours of quasars at rest-frame wavelengths of 1--2$\mu$m independent of the amount of dust extinction in the quasar host. The advantage of our sample over that of \citet{Glikman:07} is that at the higher redshifts, the hot dust emission is redshifted out of the NIR bands employed for the colour selection so the red colours can more safely be attributed to dust extinction in the quasar host.

\section{Future Work}

The current work is based on a sample selected using $\sim$1000 deg$^2$ of data from the UKIDSS-LAS. The latest data release of the UKIDSS-LAS spans $\sim$3600 deg$^2$. In addition, the ongoing VISTA Hemisphere Survey will provide an additional $\sim$10,000 deg$^2$ of high-latitude sky therefore easily allowing our sample sizes to be increased by an order of magnitude over the next five years as well as probing fainter fluxes in the $K$-band. The combination of these near infra-red surveys with mid-infrared data from the all-sky \textit{WISE} survey will enable statistically significant samples of 
reddened Type 1 AGN to be assembled. We have already embarked on
spectroscopic follow-up of these reddened AGN candidates selected from
the initial data in the new surveys using Gemini-GNIRS and this new data
will allow us to tighten our constraints on the exact obscured
fraction of AGN at the main epoch of galaxy assembly and black-hole
accretion as well as to probe in detail the dependence of this
fraction on the quasar luminosity.

\section{CONCLUSIONS}

We have extended our search for NIR selected samples of
reddened quasars to cover an area of almost 1000\,deg$^2$. Within this
area, we have identified a total of 27 very bright $K$-band sources
that also have unusually red colours. Ten of these have been presented
in Hawthorn et al. In this paper we present
spectroscopic observations of another 13 candidates. Five of these are
confirmed to be highly reddened Type 1 AGN increasing our total
reddened NIR bright AGN sample to 12.

We combine the new sample and the sample from Hawthorn
et al. in order to examine the properties of the optically obscured
luminous Type 1 AGN population. Unlike previous studies, our sample
relies on a homogenous selection at NIR wavelengths without needing
detections in the optical and/or the radio. We carefully account for
the effect of H$\alpha$ equivalent widths on the infra-red colours of our
quasars and by probing luminous sources at $z>$1, eliminate any
contribution to their colours from the host galaxy. In particular, we
reach the following conclusions:

\begin{itemize}

\item{Our sample of spectroscopically confirmed reddened quasars with
  redshifts all possess broad H$\alpha$ emission indicating that
  they are canonical Type 1 AGN.}

\item{Almost all H$\alpha$ line profiles are best fit by a broad
  ($>$5000\,kms$^{-1}$) and intermediate ($>$1500\,kms$^{-1}$)
  component and some of the lines show significant velocity offsets
  between the two components indicating the presence of strong
  outflows in the broad-line region of the AGN. Such outflows are 
  not commonly seen to affect the H$\alpha$ lines in UV-luminous quasars
  and provide evidence that the reddened quasars are observed in a phase
  when they are in the process of expelling their gas and dust.}

\item{We use the broadband colours to
  fit model SEDs to all quasars with redshifts and find that
  significant dust extinction of A$_V$=2--6 mags is required
  to fit the observational data making these quasars highly obscured
  at optical wavelengths. The dust extinctions in our sample are comparable
  to those seen in the submillimetre galaxy population at similar redshifts and 
  this level of dust obscuration pushes the quasars well below the flux limits of
  large-area optical surveys like the SDSS.}

\item{We use the line profiles to derive virial black-hole masses and
  Eddington fractions for all our reddened quasars. The mean virial
  black-hole mass is 2$\times$10$^9$M$_\odot$ and the mean Eddington
  fraction is 0.5. The reddened quasars are 
  therefore already as massive as the most massive optically selected quasars but 
  still accreting mass at a relatively high rate. We argue that they probably
  represent a short-lived phase in quasar evolution
  when the most massive starbursts are transitioning to UV-luminous quasars.}

\item{Two of our confirmed reddened quasars, ULASJ1539+0557 ($z=2.658$) and ULASJ1234+0907 ($z=2.503$)
  are intrinsically brighter than any known SDSS spectroscopic quasars
  and have mass accretion rates in excess of 100M$_\odot$yr$^{-1}$. They both lie
  at the high-redshift end of our sample consistent with a picture where the most heavily 
  reddened luminous quasars are accreting the bulk of their mass at early times via an obscured phase.}

\item{We consider the possibility of lensing in our population and
  conclude that while we cannot rule out the possibility that
  individual objects are lensed, our sample is generally fainter
  in the $K$-band than all the known lensed reddened quasars
  previously found. Crucially, many of our sources are undetected in
  the SDSS $i$, $z$ and UKIDSS $Y$ and $J$ bands where a lensing
  elliptical galaxy at $z\sim$1 should be visible. The bulk of our 
  reddened quasar population is therefore unlikely to be lensed.}

\item{The broad lines combined with the large dust extinctions
  observed in our sample strongly point to a specific phase in the
  quasar life cycle, rather than orientation, being responsible for
  the red colours. We conclude that our quasars are likely to be
  observed in a phase of coeval galaxy and black-hole
  growth when the starburst is decaying but the dust has not yet fully
  cleared preventing these sources being selected as optical quasars.}

\item{We use our sample to place new constraints of the fraction of
  obscured quasars and find tentative evidence that this is a function of
  intrinsic luminosity and dust reddening. Although a large population of modest luminosity
  heavily dust-obscured quasars is not ruled out by our data given the $K$-band flux limit 
  of our sample, other independent lines of evidence point to this not being the case. At 
  bright absolute magnitudes however, where we are reasonably complete, the obscured quasars
  seem to make up a sizeable proportion of the Type 1 quasar population suggesting that NIR
  surveys may be more complete than optical surveys at the very bright-end of the quasar luminosity function at $z\sim2$.}

\item{The large obscured fractions in the most intrinsically luminous
  quasars are consistent with these sources having a longer obscured duty
  cycle. Seen another way, the obscured phase of a quasar is probably likely 
to be associated with its most intrinsically luminous stage as it corresponds to 
a time when the black-hole is accreting mass very quickly.}

\item{Given these results and their possible implications for quasar evolution, a larger sample of $K$-band selected quasars pushing to 
fainter $K$-band magnitudes is clearly required to test the hypothesis put forward here that 
the obscured fraction of Type 1 AGN is a function of quasar luminosity.}

\end{itemize}

In conclusion, we have presented a NIR-selected sample of reddened
quasars that have been chosen to lie at 1.5$<z<$3 corresponding to the
peak of star formation and black-hole accretion in the
Universe. Unlike previous studies, we detect populations of quasars
that lie well below the SDSS detection limits as well as radio-quiet
quasars that are not present in the FIRST radio survey, thus
demonstrating the potential of new large-area infra-red surveys in
studying the dust-obscured quasar population. We find that these
luminous dusty quasars are likely being observed in a phase of coeval
galaxy and black-hole growth and are in the process of expelling their gas and dust before turning into UV-luminous quasars. Our ongoing follow-up of
these sources at long wavelengths will allow their host galaxy
properties to be constrained for the first time.

Having demonstrated the success of NIR surveys in
selecting reddened quasars, we have embarked on a spectroscopic
campaign to follow-up new sources being discovered in the VISTA
Hemisphere Survey and the Wide Infrared Survey Explorer. These new
large area datasets will enable much larger samples of dust obscured
quasars to be assembled over the coming years and push to fainter $K$-band limits thereby allowing
accurate constraints to be placed on the physical properties and
global space densities of heavily obscured broad-line quasars.

\section*{Acknowledgements}

The authors acknowledge the anonymous referee for a very constructive report that helped improve this manuscript. We thank Gordon Richards for useful discussions and Chris Lidman and
Mark Swinbank for their help with preparing the SINFONI IFU observations.  MB,
PCH and RGM acknowledge support from the STFC-funded Galaxy Formation
and Evolution programme at the Institute of Astronomy.  SA-Z
acknowledges the award of an STFC Ph.D.\ studentship.

This work is based on observations collected at the European Southern
Observatory, Paranal, Chile, (383.A- 0573(A)). This research has made
use of the NASA/IPAC Extragalactic Database (NED) which is operated by
the Jet Propulsion Laboratory, California Institute of Technology,
under contract with the National Aeronautics and Space Administration.

The sky transmission spectrum was created from data that was kindly
made available by the NSO/Kitt Peak Observatory. NSO/Kitt Peak FTS
data used here were produced by NSF/NOAO.

%\bibliographystyle{./reference/mn2e.bst}
%\bibliography{./reference/aamnem99,./reference/eros_ref}

\bibliography{}

\end{document}